\documentclass[draftcls,onecolumn,12pt]{IEEEtran}

\usepackage{amssymb}
\usepackage{amsfonts}
\usepackage{array}
\usepackage{amsmath}
\usepackage{graphicx}
\usepackage[usenames,dvipsnames]{color}
\usepackage{cite}
\usepackage{url}
\usepackage{amsthm}
\usepackage{hhline,needspace}

\theoremstyle{definition}

\theoremstyle{remark}

\newtheorem{theorem}{Theorem}

\newtheorem{proposition}{Proposition}
\newtheorem{conjecture}{Conjecture}

\usepackage{xcolor}
\newcommand{\shade}[1]{%
 \colorbox{red!20}{$\displaystyle#1$}}

\newcommand{\shadetwo}[1]{%
	\colorbox{blue!20}{$\displaystyle#1$}}

\allowdisplaybreaks

\def\AG1#1{{\textcolor{blue}{#1}}}

\DeclareMathOperator{\sinc}{sinc}

\def\Ps{P_{\rm s}}

\def\Phis{\Phi_{\rm s}}
\def\lambdas{\lambda_{\rm s}}

\def\supp{\operatorname{supp}}
\begin{document}

\title{SIR Asymptotics in General Network Models}%
\author{\IEEEauthorblockN{Anjin Guo, \emph{Student Member, IEEE}, Martin Haenggi, \emph{Fellow, IEEE},  \\
and Radha Krishna Ganti, \emph{Member, IEEE} \\ }
\thanks{Manuscript date \today. A.~Guo and M.~Haenggi are with the Dept.~of Electrical Engineering, University of Notre Dame, IN 46556, USA (e-mail: \{aguo, mhaenggi\}@nd.edu). 
R.~K.~Ganti is with the Dept.~of Electrical Engineering, Indian Institute of Technology Madras, Chennai
600036, India (e-mail: rganti@ee.iitm.ac.in).

The support of the U.S.~National Science Foundation (grant CCF 1525904) is gratefully acknowledged.}
}

\maketitle

\begin{abstract}
In the performance analyses of wireless networks, asymptotic quantities and properties often provide useful results and insights.   
The asymptotic analyses become especially important when complete analytical expressions of the performance metrics
of interest are not available, which is often the case if one departs from very specific modeling assumptions.
In this paper, we consider the asymptotics of the SIR distribution in general wireless network models, including ad hoc
and cellular networks, simple and non-simple point processes, and singular and bounded path loss models,
for which, in most cases, finding analytical expressions of the complete SIR distribution seems hopeless.  
We show that the lower tails of the  SIR distributions  decay polynomially with the order solely determined by the path loss
exponent or the fading parameter, while the upper tails decay exponentially, with the exception
of cellular networks with singular path loss. 
In addition, we analyze the impact of the nearest interferer on the asymptotic properties of the SIR distributions,
and we formulate three crisp conjectures that---if true---determine the asymptotic behavior in many cases based on the
large-scale path loss properties of the desired signal and/or nearest interferer only.

\end{abstract}


\section{Introduction}

\subsection{Motivation}
Asymptotic analyses have been widely conducted in various research areas related to wireless communications.
Although they do not quite provide the same information as complete (non-asymptotic) results,
they usually give very useful insights, while being much more tractable and available for larger classes of models.
For example,
the asymptotic coding gain in coding theory \cite{CodingGain_book} characterizes the difference of the signal-to-noise ratio (SNR) levels between the uncoded system and coded system required to reach the same bit error rate (BER) in the high-SNR regime (or equivalently,
the low-BER regime); 
the diversity gain and the multiplexing gain introduced in \cite{Diversity_multiplexing_Tse2003} are also high-SNR asymptotic metrics that
crisply capture the trade-off between the SNR exponents of the error probability and the data rate in MIMO channels;
in wireless networks, the asymptotic transmission capacity \cite{High-SIR_Transmission_Ganti2011} gives the network performance
when the density of interferers goes to 0, or equivalently, in the high signal-to-interference ratio (SIR) regime. 
These asymptotic analyses provide simple and useful results that capture important design trade-offs. 
In this paper, we focus on the  asymptotic analyses of the SIR distribution in wireless networks, which is
 a key metric that determines many other performance metrics, such as the achievable reliability, transmission rate, and the delay---it is  instrumental for the analysis and design of interference-limited wireless networks \cite{Haenggi_JSAC09}.

Our analysis is not limited to one scenario but comprehensively covers a wide range of models, including both ad hoc and cellular networks, both singular and bounded path loss laws, and general stationary point processes\footnote{This includes arbitrary superpositions of dependent or independent stationary simple point processes, as long as the superposition remains simple.}
 and general fading---unless otherwise specified. Besides, we also consider networks where each location of the transmitter (or antenna) has another transmitter (or antenna) colocated, which results in non-simple point processes.  

In all scenarios, we mainly analyze the asymptotic properties of the SIR distribution. For cellular networks with Nakagami-$m$ fading and both singular and bounded path loss models, it has been observed in \cite{GuoTCOM15, MISR2014,MISR_Ganti2015} that the success probability, defined as the  complementary cumulative distribution function (CCDF) of the SIR, for different point processes  are  horizontally shifted versions of each other (in dB) in the high-reliability regime.  Generally, in non-Poisson networks, the  success probability is  intractable. Under this observation, however, we  can obtain good approximations of the lower part of the SIR distribution (coverage probabilities above 3/4) for non-Poisson networks if we know the result of the Poisson networks and the corresponding  shift amounts.  For the tail of the SIR distribution, a similar property holds, which has been proved in \cite{MISR_Ganti2015}, if  the singular path loss model is applied. In general,
the horizontal gaps of the SIR distributions between a point process and the Poisson point process (PPP) at both ends differ slightly, so for higher accuracy, the two asymptotic regimes should be treated separately. 

This paper summarizes the known  asymptotic properties, derives  results for scenarios that have not been previously studied, and gives insight about the factors that mainly determine the behavior of the SIR.  

The reasons that we focus on the asymptotic SIR analysis include the following: 
\begin{enumerate}
\item It captures succinctly the performance of the various network models (especially for the high-reliability regime). 
\item It permits the isolation of the key network properties that affect the SIR distribution.
\item It gives insight into when it is safe to use the singular path loss model instead of a bounded one.
\item  It shows when a nearest-interferer approximation is accurate. 
\item The tail determines whether the mean SIR (and higher moments) exist.
\end{enumerate}

\subsection{Prior Work on the Analysis of the SIR Distribution}
For Poisson networks, the SIR distribution has been derived in exact analytical form in a number of cases, namely
for bipolar ad hoc networks\footnote{A bipolar network model is a model where transmitters form a stationary point process
and each transmitter has a dedicated receiver.} with general fading \cite{net:Baccelli09jsac}, for ad hoc and cellular networks with
successive interference cancellation \cite{net:Zhang14tit}, for multitier cellular networks (HetNets) with strongest-on-average
base station association and base station cooperation without retransmissions \cite{Nigam2014} and with retransmissions
\cite{net:Nigam15jsac}, for cellular networks with intra-cell diversity and/or base station silencing \cite{ICIC_ICD_Xinchen2014} and for multitier cellular networks with instantaneously-strongest base station association
\cite{net:Mukherjee12jsac,net:Blaszczyszyn15tit}. While for some specific assumptions, closed-form expressions are available,
 the results for the SIR distribution typically involve one or more integrals.
 
 The only exact result for a non-Poisson network is
 given in \cite{net:Miyoshi14aap} for cellular networks whose base stations form a Ginibre process; it contains several nested
 integrals and infinite sums and is very hard to evaluate numerically.
 
 From these exact results, simple asymptotic ones can often be derived, see, e.g., \cite{Nigam2014,ICIC_ICD_Xinchen2014},
 where results on the diversity gain are extracted from the more complicated complete results. The true power of the asymptotic
 approach, however, becomes apparent when general non-Poisson models are considered, for which essentially no exact results are available.

In general ad hoc networks modeled using the bipolar model, the asymptotic SIR distribution as the interferer density goes to 0 has been
analyzed for Rayleigh fading in \cite{TON11} and for general fading in \cite{High-SIR_Transmission_Ganti2011}.
In \cite[Ch.~5]{NOW}, the interference in general ad hoc networks has been analyzed. Relevant to our work here are the
bounds on the CCDF of the interference and the asymptotic result on the interference distribution for both 
singular and bounded path loss models.
In \cite{GuoTCOM15}, we analyzed the asymptotic properties of the signal-to-interference-plus-noise ratio (SINR) distribution in general cellular networks in the high-reliability regime for Nakagami-$m$ fading and both singular and bounded path loss models. 
In \cite{MISR2014}, a simple yet versatile analytical framework for approximating the SIR distribution in the downlink of cellular systems was proposed using the mean interference-to-signal ratio, while 
\cite{MISR_Ganti2015} considers general cellular networks with general fading and singular path loss and studies
the asymptotic behavior of the SIR distribution both at 0 and at infinity.

\subsection{Contributions} 
\label{sec::summary}
The main contribution of the paper is a comprehensive analysis of the asymptotic properties of the CCDF of the SIR
$\Ps(\theta) \triangleq \mathbb{P}({\rm SIR} > \theta)$, often referred to as the {\em success probability} or {\em coverage probability}. Regarding the transmitter/base station (BS)  distributions, we do not restrict ourselves to  simple point processes (where there is only one point at one location almost surely), which are the ones used in almost all the literature, but also consider   ``duplicated-2-point" point processes. The duplicated-2-point point processes are defined as the point processes where there are two  points at the same location, i.e., each point is duplicated. The motivation to study this model is three-fold:
(1) By comparing the results with those for standard (simple) network models, it becomes apparent whether the asymptotic behavior critically
depends on the fact that the distances to the desired transmitter and to the interferers are all different (a.s.). (2) There are situations where it is natural to consider a model where two nodes are at the same distance, such as when edge users (who have two base stations
at equal distance) in cellular networks are analyzed or when spectrum sharing between different operators is assumed and the two operators share a base station tower.
(3) The only kind of networks that are not captured by simple models are those that are formed by two fully correlated point processes.
The ``duplicated" models fills this gap.

The  asymptotic properties of  $\Ps(\theta)$ are  
summarized in Table \ref{T_total}  with respect to $\theta$ as $\theta \to 0$ and  $\theta \to \infty$ for both singular  and bounded path loss models, both ad hoc models and cellular models, and both simple point processes and duplicated-2-point point processes.   
Our results show that  the asymptotic SIR behavior is determined by two  factors---$m$ and $\delta$. $m$ is the Nakagami fading parameter  and $\delta \triangleq 2/\alpha$, where $\alpha$ is the path loss exponent.  As will be apparent from the proofs, the pre-constants
are also known in some cases, not just the scaling behavior.

The \shade{\textrm{shading}} indicates that the results have been derived in the literature marked with the corresponding reference; the marker (*) indicates that the results are only proven for the Poisson case  with Rayleigh fading, while (**) indicates that the results are only proven for the case of Rayleigh fading and the duplicated-2-point point process where the distinct locations form a PPP.

\begin{table}
	\small
	\caption{Asymptotic Properties (``Simple": simple point processes; ``Duplicated": duplicated-2-point point processes)}
	\begin{center}
		\begin{tabular}{|c|c|c|}
			\hline
			Models &$\theta \to 0$ &$\theta \to \infty$  \\ \hline 
			Simple \& Ad Hoc \& Singular path loss &$1-\Ps(\theta) = \Theta( \theta^{\delta})$ &\shade{ \Ps(\theta)  = e^{-\Theta(\theta^{\delta} )}} \cite{book} (*) \\ \hline
			Simple \& Ad Hoc \& Bounded path loss &$1-\Ps(\theta) = \Theta(  \theta^{m})$  &   $\Ps(\theta)  = e^{-\Theta(\theta^{\delta} )}$  (*) \\ \hhline{|=|=|=|}
			Simple \& Cellular \& Singular path loss  &\shade{1-\Ps(\theta) = \Theta( \theta^{m}) } \cite{GuoTCOM15}& \shade{\Ps(\theta) = \Theta(   \theta^{-\delta})} \cite{MISR_Ganti2015} \\ \hline
			Simple \& Cellular \& Bounded path loss & \shade{1-\Ps(\theta) = \Theta( \theta^{m}) } \cite{GuoTCOM15}&   $\Ps(\theta)  = e^{-\Theta(\theta^{\delta} )}$ (*)   \\ \hhline{|=|=|=|}
			Duplicated \& Ad Hoc \& Singular path loss  & $1-\Ps(\theta) = \Theta( \theta^{\delta})$ &    $\Ps(\theta)  = e^{-\Theta(\theta^{\delta} )}$ (**)  \\ \hline
			Duplicated \& Ad Hoc \& Bounded path loss &  $1-\Ps(\theta) = \Theta( \theta^{m})$   &  $\Ps(\theta)  = e^{-\Theta(\theta^{\delta} )}$ (**)  \\ \hhline{|=|=|=|} 
			Duplicated \& Cellular \& Singular path loss & $1-\Ps(\theta) = \Theta(  \theta^{m}) $  & $\Ps(\theta) = \Theta(  \theta^{-\delta-m})$ \\ \hline
			Duplicated \& Cellular \& Bounded path loss & $1-\Ps(\theta) = \Theta( \theta^{m})$   &   $\Ps(\theta)  = e^{-\Theta(\theta^{\delta} )}$ (**) \\ \hline
		\end{tabular}
	\end{center}
	\label{T_total}
\end{table}%

\begin{figure}
	\centering
	\includegraphics[width=\columnwidth]{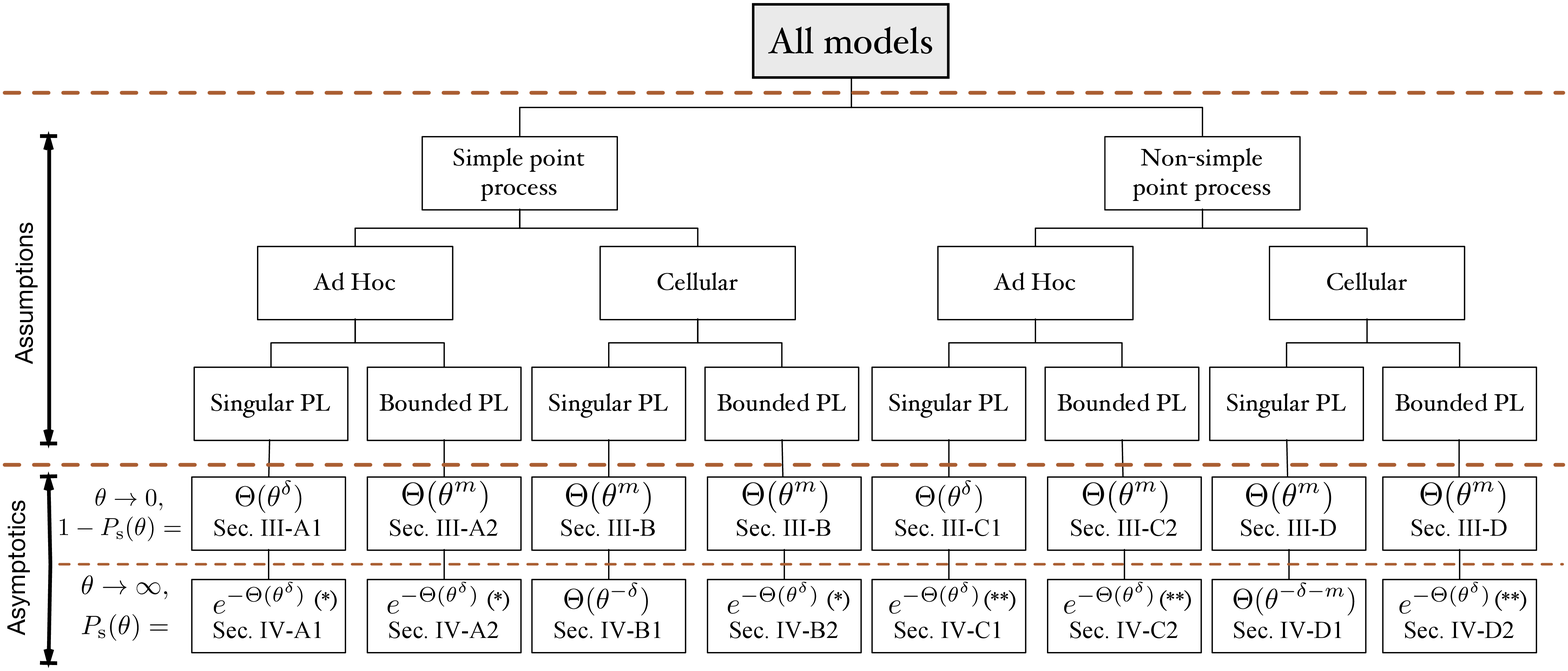}\\
	\caption{The organization of the paper sorted by the assumptions and the asymptotic results.}
	\label{Model_Structure}
\end{figure}

Alternatively, we show our results as the leaves of the tree in Fig. \ref{Model_Structure}. Visually, the tree provides the structure of the main part of this paper. We discuss the asymptotic  results for all combinations of the assumptions.  

Besides, we  study the impact of the nearest interferer on the asymptotic properties of the SIR distribution, to see whether the nearest interferer plays a dominant role in determining the asymptotics and to determine whether the nearest-interferer approximation is accurate.   

Lastly, based on the insight obtained, we offer three basic conjectures that---if they hold---capture the asymptotics of a large class
of models merely based on the large-scale path loss of the desired signal and/or the nearest interferer.

The rest of this paper is organized as follows. In Section \ref{sec::sytemmodels}, we introduce the system models. 
We analyze the asymptotic properties of the lower and upper tails of the SIR distribution in Sections \ref{sec::SIR Distribution Near 0} and \ref{sec::Tail of the SIR Distribution}, respectively. In Section \ref{sec:ImpactNearest}, the impact of the nearest interferer on the asymptotic is investigated and the three conjectures are stated.
Conclusions are drawn in Section \ref{conclusions}.

\section{System Models}
\label{sec::sytemmodels}

We consider general network models, including ad hoc networks and cellular networks, where the transmitters/BSs are assumed to follow a  stationary point process $\Phi$. Without loss of generality, we focus on the 
SIR distribution at the typical receiver at the origin $o$. We assume that the desired transmitter/base station is   $x_0$ and  all transmitters/BSs transmit at the same unit power level. 
In ad hoc networks, $x_0 $ does not belong to $ \Phi$, but in cellular networks, $x_0$ is a point of $\Phi$. Also,  in ad hoc networks, if $\Phi$ is not simple, we assume there is an interferer at the same location as $x_0$.  
Let $\Phi^{*}$ be the collection of all interferers  in both ad hoc networks and cellular networks. 
All signals experience i.i.d. fading with unit mean and the cumulative distribution function (CDF) of the fading is denoted by $F_h$.  The SIR is given by 
\begin{equation}
{\rm SIR} \triangleq \frac{S}{I} = \frac{h_{x_0} \ell(x_0) }{\int_{\mathbb{R}^2} h_x \ell(x) \Phi^{*}(\mathrm{d}x)},
\end{equation}
where $(h_x)$ are  the fading random variables and $\ell(\cdot)$ is the path loss law. We use the integral form of the interference \cite{book} instead of the usual sum  over all interferers, since $\Phi^*$ is not necessarily simple.   
Here $\Phi^*(\cdot)$ is interpreted as a counting measure.  
Note that for simple point processes, $S$ and $I$ are independent  in ad hoc networks  but correlated in cellular networks; for non-simple  point processes,  $S$ and $I$ are always correlated, since there is an interferer at the same location as the desired transmitter/BS.  

Using the notation of the interference-to-(average)-signal ratio (ISR) defined in \cite{MISR_Ganti2015}, in general,  the success probability can be expressed as
\begin{align}
\Ps(\theta) &= \mathbb{E} \bar{F}_h\left(\theta \cdot \frac{I}{\mathbb{E}_h\left( S \right)} \right) \nonumber\\
&= \mathbb{E} \bar{F}_h(\theta \cdot {\rm I\bar{S}R}), 
\end{align} 
where $\mathbb{E}_h\left( S \right) = \mathbb{E}\left( S\mid \Phi \right) =\ell(x_0)$  is the mean received signal power averaged only over the fading and $\bar{F}_h$ is the CCDF of the fading random variables.

In the following two sections, we will first discuss the asymptotic properties of the lower tail of the SIR distribution (near 0) and then the upper tail of   the SIR distribution (near $\infty$). Note that in the rest of this paper, by ``tail" we mean the ``upper tail", whereas ``lower tail" refers to the asymptotics near 0.  

\section{Lower Tail of the SIR Distribution} 
\label{sec::SIR Distribution Near 0}
\subsection{Simple Ad Hoc Models}

\subsubsection{Singular path loss model}
\label{sec:simple_singular_0}
Consider a wireless network where all transmitters follow a simple stationary point processes $\Phi$ of intensity $\lambda$ and  the distance between the transmitter and the corresponding receiver is a constant $b>0$. 
We add an additional transmitter-receiver pair with the  receiver at the origin $o$ and its desired transmitter at $x_0 = (b,0) \notin \Phi$ and analyze the SIR distribution at $o$. 
We assume all transmitters are always transmitting\footnote{This is not a restriction due to the generality of the point process model (most MAC schemes preserve the stationarity of the transmitters).} at unit power and in the same frequency band. 
Every signal is assumed to experience i.i.d.  fading with mean 1.  The path loss model is $\ell(x) = \|x\|^{-\alpha}$, where $\alpha > 2$.  
The lower tail of the CDF of the SIR has the following property: 
\begin{theorem} Let $\Phi$ be a simple stationary  point processes   with intensity $\lambda$, 
	and let   the desired received  signal strength be given by $S= h b^{-\alpha}$ and the interference be given by  $I = \sum_{x \in \Phi} h_x \|x\|^{-\alpha}$. 
	If the fading random variable $h$ with mean 1 satisfies    
	$\mathbb{E}[h^{-\delta}] < \infty$,  
	 we have 
	\begin{equation}
	1-\Ps(\theta) \sim \theta^{\delta}   \pi \lambda b^2 \mathbb{E}[h^{\delta}] \mathbb{E}[h^{-\delta}], \quad \theta \to 0. 
	\label{asymp_simple_singular_0}
	\end{equation}
	In particular, if $h\sim {\rm exp}(1)$, $1-\Ps(\theta) \sim \theta^{\delta}   \pi \lambda b^2 \Gamma(1+\delta)\Gamma(1-\delta)$, as $\theta \to 0$.
	\label{Thm_simple_singular_0}
\end{theorem}
\begin{IEEEproof}
	Using the same method as in the proof of Theorem 5.6 in \cite{NOW}, we can show that 
	\begin{equation}
	\mathbb{P}(I\geq y) \sim \pi \lambda \mathbb{E}[h^{\delta}] y^{-\delta}, \quad y \to \infty.
	\label{heavy_tail_I}
	\end{equation}
	Note that in the proof of Theorem 5.6 in \cite{NOW}, the reduced Palm measure is used. In our case, we use the standard probability measure, since in our model, the  transmitter and receiver do not belong to $\Phi$, while in Theorem 5.6 in \cite{NOW}, the result is conditioned on that there is a transmitter belonging to $\Phi$.   
	The success probability can be rewritten as
	\begin{align}
	\Ps(\theta) &= 1- \mathbb{P}\left(    I >h_{x_0} b^{-\alpha}  \theta^{-1} \right)  \nonumber\\
	&= 1- \mathbb{E}_{h}\left[ \mathbb{P}\left(    I >h b^{-\alpha}  \theta^{-1} \mid h  \right) \right].   
	\end{align}
	Thus, 
	\begin{align}
	\lim_{\theta \to 0} \frac{1- \Ps(\theta)}{\theta^{\delta}} &=\lim_{\theta \to 0} \frac{ \mathbb{E}_{h}\left[ \mathbb{P}\left(    I >h b^{-\alpha}  \theta^{-1} \mid h  \right) \right]}{\theta^{\delta}} \nonumber\\
	&\stackrel{(a)}{=}    \mathbb{E}_{h}\left[  \lim_{\theta \to 0} \frac{ \mathbb{P}\left(    I >h b^{-\alpha}  \theta^{-1} \mid h  \right)}{\theta^{\delta}}  \right] \nonumber\\ 
	&\stackrel{(b)}{=}     \pi \lambda b^2 \mathbb{E}[h^{\delta}] \mathbb{E}[h^{-\delta}],  
	\end{align}
	where $(a)$ follows from the dominated convergence theorem and $(b)$ follows from \eqref{heavy_tail_I}. 
\end{IEEEproof}

All standard fading models, such as Nakagami-$m$ fading, Rician fading and lognormal fading,  satisfy the conditions in Theorem \ref{Thm_simple_singular_0}. 

In this model, the distance between the nearest interferer and the origin could be arbitrarily small irrespective of the type of the point process and thus the ratio of  the  average desired signal strength and the  nearest interferer's signal strength averaged over the fading could be arbitrarily small due to the singular path loss.  
In the rest of this subsection, we study whether the nearest interferer determines the asymptotic property of $\Ps(\theta)$, as $\theta \to 0$.  
The following proposition adapts the result from \cite[Lemma 7]{MISR_Ganti2015}\footnote{In Lemma 7 of \cite{MISR_Ganti2015}, it is shown that the tail of the CCDF of the desired signal strength $S$ in cellular networks where each user is associated with its nearest BS is $\mathbb{P}(S>y) \sim \lambda \pi \mathbb{E}\left[h^{\delta}   \right] y^{-\delta}$, as $y \to \infty$. The signal strength of the nearest BS in \cite{MISR_Ganti2015} corresponds to that of the nearest interferer in our simple ad hoc models, since they have the same distribution.} to the ad hoc model and gives the property of the upper tail of the CCDF of the nearest interferer's signal strength. 
\begin{proposition}
	\label{lemma1}
	For all stationary point processes, the tail of the CCDF of the nearest interferer's  signal strength $I_0$ at the  receiver $o$, i.e., $I_0 \triangleq h \cdot \max_{x\in \Phi^*} \ell(x)$,  is 
	\begin{equation}
	\mathbb{P}(I_0>y) \sim \lambda \pi \mathbb{E}[h^{\delta}] y^{-\delta}, \quad y \to \infty.
	\end{equation}
\end{proposition}

Proposition \ref{lemma1}   implies that the CDF of the ratio of  the  desired signal strength and the nearest interferer's signal strength, denoted by $\frac{S}{I_0}$, satisfies
\begin{align}
\mathbb{P}\left(\frac{S}{I_0}<\theta \right) &= P( I_0 > h b^{-\alpha} \theta^{-1} )  \nonumber\\
	&\sim \lambda \pi  \mathbb{E}(h^\delta) \mathbb{E}_h\left[\left(h b^{-\alpha} \theta^{-1}\right)^{-\delta}\right]  \nonumber\\
	&= \lambda \pi b^2  \mathbb{E}[h^\delta]\mathbb{E}[h^{-\delta}] \theta^{\delta}, \quad \theta \to 0. 
\label{SNearestIR}
\end{align} 

So,  $\mathbb{P}\left({\rm SIR}<\theta \right) > \mathbb{P}\left(\frac{S}{I_0}<\theta \right) \sim \lambda \pi b^2 \mathbb{E}\left[h^{\delta}   \right] \mathbb{E}\left[{h}^{-\delta}\right]  \theta^{\delta}$, as $\theta \to 0$.   

By Theorem \ref{Thm_simple_singular_0}, we find  that $\mathbb{P}\left({\rm SIR} <\theta \right)\sim  \mathbb{P}\left(\frac{S}{I_0}<\theta \right)$, as $\theta \to 0$. So the nearest interferer alone determines the asymptotic behavior of $\Ps(\theta)$ (not just the pre-constant), as $\theta \to 0$. For Nakagami-$m$ fading, $\mathbb{E}[h^{\delta}]\mathbb{E}[h^{-\delta}] = \frac{\Gamma(m)^2}{\Gamma(m+\delta)\Gamma(m-\delta)}$. So, both $m$ and $\delta$ affect the pre-constant, but only $\delta$ determines the decay order.

\subsubsection{Bounded path loss model}
We assume that the path loss model 
is $\ell(x) = (\epsilon+\|x\|^{\alpha})^{-1}$, where $\alpha > 2$ and $\epsilon>0$, and that signals experience Nakagami-$m$ fading with mean 1, i.e.,    
the fading variables are distributed as  $h \sim \textrm{gamma}(m,\frac{1}{m})$. We have 
\begin{equation}
\lim_{t \to 0} \frac{F_h(t)}{t^m}  = \lim_{t \to 0} \frac{(mt)^{m-1}\exp(-mt)}{\Gamma(m) t^{m-1}}=\frac{m^{m-1}}{\Gamma(m)}. 
\end{equation}

So as $\theta \to 0$, 
\begin{align}
1-\Ps(\theta) 
&=   \mathbb{P}\left( \frac{h_0 (\epsilon+b^{\alpha})^{-1}}{I} < \theta \right) \nonumber \\
&=   \mathbb{E} \left[ {F}_{h}\left( \theta (\epsilon+ b^{\alpha})   I  \right)\right] \nonumber \\
&\stackrel{(a)}{\sim} \theta^{m} \frac{m^{m-1}}{\Gamma(m)} (\epsilon+ b^{\alpha})^m \mathbb{E}[I^m], \quad \epsilon>0,
\end{align}
where   $I = \sum_{x\in \Phi } h (\epsilon+\|x\|^{\alpha})^{-1}$ and $(a)$ follows by the dominated convergence theorem (similar to the proof of Theorem 1 in   \cite{GuoTCOM15}). By  Lemma 1 in \cite{GuoTCOM15}, we have $\mathbb{E}[I^m] < \infty$, for $\epsilon>0$.

\subsection{Simple Cellular Models}
For both singular and bounded path loss models with Nakagami-$m$ fading, the results in Table \ref{T_total} have been proved in  \cite{GuoTCOM15}.

\subsection{Non-simple Ad Hoc Models}

\subsubsection{Singular path loss model}
\label{sec::asymp_nonsimple_singular_0} 
We assume that in the wireless network, all transmitters follow a duplicated-2-point stationary point processes $\Phi$ of intensity $\lambda$, where every point has a partner point colocated. We add an additional transmitter-receiver pair with the receiver at the origin $o$ and its desired transmitter at $x_0 = (b,0) \notin \Phi$ and an additional transmitter $x_1 = x_0$ as an interferer. We analyze the SIR distribution at $o$.
As  in Section \ref{sec:simple_singular_0}, we assume all transmitters are always transmitting at unit power and  in the same frequency band; the path loss model is $\ell(x) = \|x\|^{-\alpha}$, where $\alpha > 2$; every signal is assumed to experience i.i.d. fading with mean 1 and PDF $f_h$.

$\Phi$ can no longer be represented as a random set, since a set can only contain one instance of each element.  Let  $\Phis$ be the simple point process version of $\Phi$, which means that at every point location of $\Phi$, there is only one point of $\Phis$. So $\Phis$ is a random set. 
Viewing $\Phi$ and $\Phis$ as random counting measures, we thus have $\Phi=2\Phis$.  
The intensity of $\Phis$  is $\lambdas = \lambda/2$.  For the case where $\Phis$ is a PPP, the SIR distribution for Rayleigh fading
has been obtained as a limiting case of the Gauss-Poisson process as the distance between the two points forming a cluster
goes to zero \cite{net:Guo16tcom}. Here we consider more general models, both for $\Phis$ and for the fading.

Let $\tilde{I} \triangleq  \sum_{x \in \Phis} \left(h_{x,1}+h_{x,2}\right) \ell(x)$, where $h_{x,1}$ and $h_{x,2}$ are the two fading variables of the transmitters at  the location of $x$.    The total interference, including the one from the partner node of the desired transmitter, is then given by $I = h_1 b^{-\alpha} + \tilde{I}$. 
The SIR at the  receiver at $o$ can be expressed as 
\begin{equation}
{\rm SIR} = \frac{h_0 b^{-\alpha}}{h_1 b^{-\alpha} + \tilde{I}}, 
\end{equation}
where $\{h_0, h_1\}$ are fading variables. 

The following theorem characterizes  the lower tail of the   SIR distribution. 
\begin{theorem} Let $\Phi$ be a stationary point process with intensity $\lambda$, where  every transmitter is colocated with another one.  
	We focus on a  receiver at  the origin $o$, with the desired received  signal strength $S= h_{0}b^{-\alpha}$ and the interference $I =\sum_{x \in \Phis} \left(h_{x,1}+h_{x,2}\right) \|x\|^{-\alpha} + h_{1}b^{-\alpha}$.  
	If the fading random variables $ h_a, h_b $ satisfy   
	$\mathbb{E}[h_a^{-\delta}] < \infty$ and $\mathbb{P}\left(h_a <  \theta h_b \right) = \Theta(\theta^{K})$ as $\theta \to 0$, 
	where $K\geq 1$,  
	then we have 
	\begin{equation}
	1-\Ps(\theta) \sim \theta^{\delta}  \frac{ \pi \lambda b^2}{2} \mathbb{E}[\left(h_a+h_b\right)^{\delta}] \mathbb{E}[h^{-\delta}], \quad \theta \to 0, 
	\label{asymp_nonsimple_singular_0}
	\end{equation}
	where $h, h_a, h_b$  
	are iid  fading random variables with mean 1. 
	Specifically, Nakagami-$m$ fading meets the fading constraints with $K=m$.
	\label{Thm_nonsimple_singular_0}
\end{theorem}
\begin{IEEEproof}
	See Appendix \ref{sec:app_nonsimple_singular_0}. 
\end{IEEEproof}

Comparing \eqref{asymp_nonsimple_singular_0} with \eqref{asymp_simple_singular_0}, we observe that  by duplication of the transmitters, only the pre-constant changes, since for the interference, the duplication can be interpreted as the duplication of the fading variable and the decay order is determined only by $\delta$.

\subsubsection{Bounded path loss model}
\label{sec::asymp_nonsimple_bounded_0}
The system model  is the same as that in Section \ref{sec::asymp_nonsimple_singular_0}, except that  $\ell(x) = (\epsilon+\|x\|^{\alpha})^{-1}$, where $\alpha > 2$ and  $\epsilon>0$, and that signals experience Nakagami-$m$ fading with mean 1.

As $\theta \to 0$, we have 
\begin{align}
1-\Ps(\theta) 
&=   \mathbb{P}\left( \frac{h_0 (\epsilon+b^{\alpha})^{-1}}{h_1 (\epsilon+b^{\alpha})^{-1}+ \tilde{I}} < \theta \right) \nonumber \\
&=   \mathbb{E} \left[ {F}_{h}\left( \theta (\epsilon+ b^{\alpha})  \left( h_1 (\epsilon+b^{\alpha})^{-1}+ \tilde{I} \right)  \right)\right] \nonumber \\
&\stackrel{(a)}{\sim} \theta^{m} \frac{m^{m-1}}{\Gamma(m)} (\epsilon+ b^{\alpha})^m \mathbb{E}\left[\left( h_1 (\epsilon+b^{\alpha})^{-1}+ \tilde{I} \right)^m \right], 
\end{align}
where $(a)$ follows by the dominated convergence theorem (similar to the proof of Theorem 1 in \cite{GuoTCOM15}). By  Lemma 1 in \cite{GuoTCOM15}, we have $\mathbb{E}[\tilde{I}^n] < \infty$ for any $n \in \mathbb{N}$, and thus $\mathbb{E}\left[ \left( h_1 (\epsilon+b^{\alpha})^{-1}+ \tilde{I} \right)^m \right]<\infty$ follows using binomial expansion.

\subsection{Non-simple Cellular Models}
\label{sec::asymp_nonsimple_singular_0_cellular}

Consider a downlink cellular network model. The base station (BS) locations are modeled as a stationary point process $\Phi$ with intensity $\lambda$, where every point has a partner point colocated. Without loss of generality, we assume that the typical user is located at the origin $o$ and is served by one of the two  nearest BSs.
All transmitters are assumed to be always transmitting signals using unit power and in the same frequency band. 
Every signal is assumed to experience i.i.d. Nakagami-$m$ fading with mean 1 and PDF $f_h$, and  there is no noise. 

For both the singular path loss model $\ell(x) = \|x\|^{-\alpha}$ and the bounded path loss model $\ell(x) = \left(\epsilon+ \|x\|^{\alpha}\right)^{-1}$, where $\alpha > 2$,  
we can  simply modify the proof of Theorem 1 in \cite{GuoTCOM15} and  prove that  $1-\Ps(\theta) \sim \theta^{m}$, as $\theta \to 0$. 

As pointed out earlier, 
this model can be applied to the analysis of edge users of two adjacent cells in cellular networks, since each edge user has an interferer that has the same distance to the user as the serving BS (assuming frequency reuse 1), or to analyze the benefits of spectrum sharing
between cellular operators who share base station towers.
These applications' model is slightly  different from our model, since the interferers, excluding the nearest one,  do not have partner points colocated. However, as we shall see, the scaling behavior of the SIR remains the same, irrespective of whether all points are duplicated
or only the nearest one.

\subsection{Discussion}
We use the following terminology for the discussion: A tail with $\theta^{\pm\delta}$ (where $\delta<1$) is said to follow a 
{\em heavy power law}\footnote{The positive exponent applies to $\theta\to 0$, while the negative exponent applies to $\theta\to\infty$.},
one with $\theta^{\pm m}$ with $m\geq 1$ follows a {\em light power law}, while one with 
$e^{-\Theta(\theta^\delta)}$ is {\em exponential}.

In Table \ref{T_total}, for those entries where the lower or upper tail of the SIR distribution follows a heavy power law, the results are true for essentially all motion-invariant (m.i.) point processes. So the decay order is a function of  $\delta$ if either
the distribution of $\|x_0\|$ or the distributions of the distances from interferers to the origin determine the decay order. 

For non-trivial asymptotics of $\mathbb{P}(S/I<\theta)$ as $\theta\to 0$, we need $0\in\supp(S)\cup\supp(1/I)$, where $\supp(X)$ is
the support of the continuous random variable $X$.
If this is not the case, there is a $\theta_0>0$ s.t.~$\Ps(\theta_0)=1$ $\forall \theta<\theta_0$, and thus the lower tail is trivial.

In ad hoc models, for both the singular and bounded path loss models, whether $0\in\supp(S)$ is only determined by the lower tail of the  fading distribution and whether $0\in\supp(1/I)$ is determined by both the tail of the fading distribution and the distributions of the interferers.
We proved that  for ad hoc models with bounded path loss,  $1-\Ps(\theta)$ follows a light power law with a decay order determined by the fading parameter $m$.  For  ad hoc models with singular path loss, we proved that  the lower tail of the distributions of the interferer distances, rather than the fading distribution, dominates the decay order and thus the decay order is only determined by $\delta$.

\section{Tail of the SIR Distribution}
\label{sec::Tail of the SIR Distribution}

\subsection{Simple Ad Hoc Models}

\subsubsection{Singular path loss model}
\label{sec:simple_singular_infty}
We first consider the PPP   with Rayleigh fading and then discuss the  case of the PPP with Nakagami-$m$  fading. 
The success probability $P_{s,{\rm PPP}}(\theta)$ for the PPP is \cite[Ch.~5.2]{book}  
\begin{equation}
P_{s,{\rm PPP}}(\theta) =  \exp\left(  - \pi \lambda \theta^{\delta} b^2 \Gamma(1+\delta)\Gamma(1-\delta)   \right). 
\label{Singular_infty_Ps_PPP}
\end{equation}

Hence,   $P_{s,{\rm PPP}}(\theta)  = e^{-\Theta(\theta^{\delta} )}$, as $\theta \to \infty$.   

Note in \eqref{Singular_infty_Ps_PPP}, $P_{s,{\rm PPP}}(\theta)$ is in the form of the void probability of a ball with radius $d_0 = \theta^{\delta/2}b (\Gamma(1+\delta)\Gamma(1-\delta) )^{1/2}$, i.e., the probability that there is no interferer with distance less than $d_0$ to $o$. 

For $h \sim {\rm exp}(1)$, we have $\mathbb{P}(h>\beta) = \exp(-\beta)$. 
From the above result, 
we observe that  $P_{s,{\rm PPP}}(\theta)=\mathbb{P}\left(  h  > \theta b^{\alpha} I \right)$ does not have an exponential tail 
$e^{-\Theta(\theta)}$  
as the fading random variable does. 
Hence it is the  interference term in the denominator that determines the power of $\theta$.  

For the simple PPP case with Nakagami-$m$ fading, we have the following proposition. 
\begin{proposition}
	For the PPP $\Phi$ with intensity $\lambda$ with desired received signal strength $S= h b^{-\alpha}$ and
	interference $I = \sum_{x \in \Phi} h_x \|x\|^{-\alpha}$, where $(h_x)$ are i.i.d.~Nakagami-$m$ fading variables,  
we have 
	\begin{equation}
	P_{s,{\rm PPP}}(\theta) = e^{-\Theta(\theta^{\delta})},  \quad \theta \to \infty. 
	\label{Adhoc_singular_PPP_nakagami_eq}
	\end{equation}
	
	\label{Adhoc_singular_PPP_nakagami}	
\end{proposition}
\begin{IEEEproof}
By \eqref{Singular_infty_Ps_PPP}, it has been showed that \eqref{Adhoc_singular_PPP_nakagami_eq} holds for $m=1$.
First, we consider the case with $m = 2$.  
The Laplace transform of the interference $I$  for the PPP is 
$\mathcal{L}_I(s) = \mathbb{E} [e^{-sI}]= \exp\left( -\pi \lambda \mathbb{E}(h^{\delta} ) \Gamma(1-\delta) s^{\delta} \right)$ \cite[Ch.~5.2]{book}.  By taking the derivative of $\mathcal{L}_I(s)$, we obtain  
\begin{align}
\mathbb{E} [Ie^{-sI}] &=\pi \lambda \mathbb{E}(h^{\delta} ) \Gamma(1-\delta) \delta s^{\delta-1} \exp\left( -\pi \lambda \mathbb{E}(h^{\delta} ) \Gamma(1-\delta) s^{\delta} \right).
\end{align}  
Using the expressions of $\mathbb{E} [e^{-sI}]$ and $\mathbb{E} [Ie^{-sI}]$ above, the success probability is expressed as 
\begin{align}
P_{s,{\rm PPP}}(\theta)  &= \mathbb{P} \left(   h   > \theta b^{\alpha} I \right) \nonumber\\
				&= \mathbb{E}_{I}\left[   \int_{\theta b^{\alpha} I }^{\infty}  4x e^{-2x} \mathrm{d} x \right]  \nonumber\\
				&\stackrel{(a)}{=}   \mathbb{E}_{I}\left[   2 \theta b^{\alpha} I e^{- 2 \theta b^{\alpha} I}  +  \int_{2\theta b^{\alpha} I }^{\infty}  e^{-x} \mathrm{d} x  \right]      \nonumber\\
				&= 2 \theta b^{\alpha} \mathbb{E}_{I}\left[   I e^{- 2 \theta b^{\alpha} I} \right] +   \mathbb{E}_{I}\left[   e^{- 2 \theta b^{\alpha} I} \right]     \nonumber\\
				&= \left( \pi \lambda \mathbb{E}(h^{\delta} ) \Gamma(1-\delta) \delta 2^{\delta} b^2 \theta^{\delta} +1 \right) \exp\left( -\pi \lambda \mathbb{E}(h^{\delta} ) \Gamma(1-\delta) 2^{\delta} b^2 \theta^{\delta}  \right), 
\end{align}
where $(a)$ follows using  integration by parts. Therefore, as $\theta \to \infty$, $-\log\left(P_{s,{\rm PPP}}(\theta)\right) = \Theta(\theta^{\delta})$. 

For $m\geq 3$, we can obtain the same result by the same reasoning as for $m=2$, i.e., applying the $(m-1)$-th  derivative of $\mathcal{L}_I(s)$ and   integration by parts. 
\end{IEEEproof}

\subsubsection{Bounded path loss model}
We consider the PPP case with Nakagami-$m$ fading and assume that the path loss model is $\ell(x) = (\epsilon+\|x\|^{\alpha})^{-1}$, where $\alpha > 2$ and $\epsilon > 0$. 

For $m = 1$, the success probability $P_{s,{\rm PPP}}(\theta)$ can be expressed as
\begin{align}
P_{s,{\rm PPP}}(\theta) =  \mathbb{E}\left[ \exp\left( - \left(\epsilon+b^{\alpha}\right) \theta I \right) \right],  
\end{align}
which is in the form of the Laplace transform of the interference $I$. 
We have 
\begin{align}
\mathcal{L}_{I}(s) &\triangleq   \mathbb{E}\left[ \exp\left( - s I \right) \right] \nonumber \\
&= \exp\left(- \lambda \int_{\mathbb{R}^2} \left(1-  \mathbb{E}_h \left[ \exp\left( - s h \left(\epsilon+\|x\|^{\alpha} \right)^{-1} \right) \right] \right) {\rm d}x \right) \nonumber \\
&= \exp\left(- 2\pi \lambda \int_{0}^{\infty} \left(1- \frac{1}{1+s  \left(\epsilon+r^{\alpha} \right)^{-1} } \right)r {\rm d}r \right) \nonumber \\
&\stackrel{(a)}{=} \exp\left(- 2\pi \lambda s (s+\epsilon)^{\delta -1} \int_{0}^{\infty}  \frac{r}{1+ r^{\alpha} } {\rm d}r \right)   \nonumber \\
&= \exp\left(- \frac{\pi \lambda}{\sinc \delta} s (s+\epsilon)^{\delta -1}   \right), 
\label{result_LIa}			 
\end{align}
where $(a)$ follows by using the substitution $r(s+\epsilon)^{-\delta/2} \to r$. 

Thus, 
$P_{s,{\rm PPP}}(\theta) =   \exp\left(- \frac{\pi \lambda}{\sinc \delta} \left(\epsilon+b^{\alpha}\right) \theta (\left(\epsilon+b^{\alpha}\right) \theta+\epsilon)^{\delta -1} \right)$. We have 
\begin{equation}
\log\left( P_{s,{\rm PPP}}(\theta) \right)  \sim  - A \theta^{\delta}, \quad \theta \to \infty,  
\label{trendm1numerical}
\end{equation} 
where $A = \frac{\pi \lambda}{\sinc \delta}  \left(\epsilon+b^{\alpha}\right)^{\delta}$. 

For $m \in \mathbb{N}$, we have the following proposition. 

\begin{proposition}
	For the PPP $\Phi$ with intensity $\lambda$ with  
	  desired received  signal strength  $S= h  (\epsilon+b^{\alpha})^{-1}$ and  interference $I = \sum_{x \in \Phi} h_x  (\epsilon+\|x\|^{\alpha})^{-1}$, where $\{h_x\}$ are i.i.d. Nakagami-$m$ fading variables,  
 we have 
	\begin{equation}
	P_{s,{\rm PPP}}(\theta) = e^{-\Theta(\theta^{\delta})},  \quad  \theta \to \infty. 
	\label{Adhoc_bounded_PPP_nakagami_eq}
	\end{equation}
	\label{Adhoc_bounded_PPP_nakagami}	
\end{proposition}
\begin{IEEEproof}
The proof is similar to that of Proposition \ref{Adhoc_singular_PPP_nakagami}. 
\end{IEEEproof}

\subsection{Simple Cellular Models}

\subsubsection{Singular path loss model}
For  general fading, the SIR asymptotics, derived in \cite{MISR_Ganti2015}, are  
\begin{align}
P_s(\theta) \sim \pi \lambda \theta^{-\delta} \mathbb{E}_o^! (I_{\infty}^{-\delta})  \mathbb{E}(h^\delta) , \quad \theta \to \infty, 
\end{align}
where $I_{\infty} \triangleq \sum_{x\in \Phi} h_x \ell(x)$. 

\subsubsection{Bounded path loss model}
\label{sec::asymp_simple_bounded_infty_cellular}
By the same reasoning as in the proof of Theorem 4 in \cite{MISR_Ganti2015}, we have 
\begin{align}
P_s(\theta) \sim \lambda \theta^{-\delta}\int_{\mathbb{R}^2} \mathbb{E}_o^! \left[ 
\bar{F}_{h}\left(  \left(\theta \epsilon+ \|x\|^{\alpha}\right) I_{\infty} \right)   \right]  \mathrm{d}x, \quad \theta \to \infty. 
\end{align}

We consider the PPP case with Nakagami-$m$ fading. For $m = 1$, as $\theta \to \infty$,  
\begin{align}
P_{s,{\rm PPP}}(\theta) &\sim \lambda \theta^{-\delta}\int_{\mathbb{R}^2} \mathbb{E}_o^! \left[ 
\exp\left(  -\left(\theta \epsilon+ \|x\|^{\alpha}\right) I_{\infty} \right)   \right]  \mathrm{d}x \nonumber\\
&\stackrel{(a)}{=}\lambda \theta^{-\delta}\int_{\mathbb{R}^2} \exp\left(- 2\pi \lambda \left(\theta \epsilon+ \|x\|^{\alpha}\right)  ( \theta \epsilon+ \|x\|^{\alpha}  +1)^{\delta -1} \int_{0}^{\infty}  \frac{r}{1+ r^{\alpha} } {\rm d}r \right)  \mathrm{d}x  \nonumber\\
&=\lambda \theta^{-\delta}   \exp\left(- \frac{\pi \lambda}{\sinc \delta}  \theta^{\delta} \epsilon^{\delta}  \right) \!
\int\limits_{\mathbb{R}^2} \exp\left(- \frac{\pi \lambda}{\sinc \delta}  \left( \left(\theta \epsilon\!+\! \|x\|^{\alpha}\right)  (\theta \epsilon \!+ \!\|x\|^{\alpha}\! +\!1)^{\delta -1} \!-\!\theta^{\delta} \epsilon^{\delta} \right)  \right) \! \mathrm{d}x, 
\end{align}
where $(a)$ follows by using the result in \eqref{result_LIa}. 

Since $\log\left(  \int_{\mathbb{R}^2} \exp\left(- \frac{\pi \lambda}{\sinc \delta}  \left( \left(\theta \epsilon+ \|x\|^{\alpha}\right)  (\theta \epsilon+ \|x\|^{\alpha} +1)^{\delta -1} -\theta^{\delta} \epsilon^{\delta} \right) \right)  \mathrm{d}x\right) = o(\theta^{\delta})$,  as $\theta \to \infty$, 
we have $\Ps(\theta)  = e^{-\Theta(\theta^{\delta} )}$, as $\theta \to \infty$.

\subsection{Non-simple Ad Hoc Models}

\subsubsection{Singular path loss model}
The system model  is the same as that in Section \ref{sec::asymp_nonsimple_singular_0}, except that $\Phis$ is assumed to be a uniform (and thus simple) PPP with intensity $\lambda/2$ and the fading is Rayleigh. 
The success probability $P_{s,{\rm PPP}}(\theta)$  is 
\begin{align}
P_{s,{\rm PPP}}(\theta) &= \mathbb{P}\left(    \frac{h_0 b^{-\alpha}}{h_1 b^{-\alpha} + \tilde{I}} > \theta  \right) \nonumber\\
&= \mathbb{E}\left[e^{-h_1 \theta} \right] \mathbb{E}\left[  e^{-\theta b^{\alpha} \tilde{I}} \right]     \nonumber\\
&=\frac{1}{1+\theta} \exp\left(  -\frac{\pi \lambda}{2} \theta^{\delta} b^2 \mathbb{E}\left[ \left( h_a+h_b \right)^{\delta}\right]  \Gamma(1-\delta)   \right), 
\label{PPP_infty_nonsimple_singuilar111}
\end{align}
where $h_0,h_1,h_a,h_b$ are iid fading variables with unit mean. Taking the logarithm on both sides, we have $-\log(P_{s,{\rm PPP}}(\theta)) -\log(1+\theta) \sim \Theta(\theta^{\delta})$, as  $\theta \to \infty$.  
Thus,  $\Ps(\theta)  = e^{-\Theta(\theta^{\delta} )}$, as $\theta \to \infty$.

Note that if we had $k$ colocated interferers ($k+1$ colocated nodes), the prefactor on the right-hand side of \eqref{PPP_infty_nonsimple_singuilar111} would be $(1+\theta)^{-k}$. When $k = 2$, the prefactor is the same as in the result for cellular worst-case users \cite[Eq. (19)]{Nigam2014}, which are users equidistant from three base stations.

\subsubsection{Bounded path loss model}
\label{sec::asymp_nonsimple_bounded_infty}
The system model  is the same as that in Section \ref{sec::asymp_nonsimple_bounded_0}, except that $\Phis$ is assumed to be a uniform PPP with intensity $\lambda/2$ and the fading is Rayleigh. 

The success probability $P_{s,{\rm PPP}}(\theta)$  is 
\begin{align}
P_{s,{\rm PPP}}(\theta) &= \mathbb{P}\left(    \frac{h_0  \left(\epsilon+b^{\alpha} \right)^{-1} }{h_1  \left(\epsilon+b^{\alpha} \right)^{-1}  + \tilde{I}} > \theta  \right) \nonumber\\
&= \mathbb{E}\left[e^{-h_1 \theta} \right] \mathbb{E}\left[  e^{-\theta \left(\epsilon+b^{\alpha} \right) \tilde{I}} \right]     \nonumber\\
&=\frac{1}{1+\theta} \exp\left(  -\pi \lambda \int_0^{\infty} \left(   1- \mathbb{E}\left[  e^{- g_2 s_1  \left(\epsilon+r^{\alpha} \right)^{-1} }\right] \right) r \mathrm{d}r  \right)  \nonumber\\
&=\frac{1}{1+\theta} \exp\left(  -\pi \lambda \int_0^{\infty} \left(   1- \frac{1}{\left(  1+s_1  \left(\epsilon+r^{\alpha} \right)^{-1}  \right)^2}  \right) r \mathrm{d}r  \right)   \nonumber\\
&\stackrel{(a)}{=} \frac{1}{1+\theta} \exp\left(  -2 \pi \lambda s_1 (s_1+\epsilon)^{\delta-1}\int_0^{\infty} \frac{r^{\alpha}+\frac{s_1+2\epsilon}{2(s_1+\epsilon)} }{\left(  r^{\alpha} + 1 \right)^2} r\mathrm{d}r  \right) \nonumber\\
&\stackrel{(b)}{\sim}  \frac{1}{1+\theta} \exp\left(  -2 \pi \lambda s_1 (s_1+\epsilon)^{\delta-1}\int_0^{\infty} \frac{r^{\alpha}+\frac{1}{2} }{\left(  r^{\alpha} + 1 \right)^2} r\mathrm{d}r  \right) , \quad \theta \to \infty   \nonumber\\
&\sim \frac{1}{1+\theta} \exp\left(  -\frac{ \pi \lambda(\delta+1)}{2 \sinc \delta} s_1 (s_1+\epsilon)^{\delta-1}  \right) , \quad \theta \to \infty,  
\end{align}
where $h_0,h_1,h_a,h_b$ are i.i.d.~fading variables, $s_1 \triangleq \theta \left(\epsilon+b^{\alpha} \right)$, $g_2  \triangleq h_a+h_2 \sim {\rm gamma}(2,1)$, $(a)$ follows from the substitution $r(s_1+\epsilon)^{-\delta/2} \to r$, and $(b)$ follows by the dominated convergence theorem.   
Thus,  $\Ps(\theta)  = e^{-\Theta(\theta^{\delta} )}$, as $\theta \to \infty$.

\subsection{Non-simple Cellular Models}

\subsubsection{Singular path loss model}
\label{sec::asymp_nonsimple_singular_infty_cellular}

The system model  is the same as that in Section \ref{sec::asymp_nonsimple_singular_0_cellular}, except that $\ell(x) = \|x\|^{-\alpha}$, where $\alpha > 2$. 

Let $x_0$ denote the serving BS of the typical user, and let $x_1=x_0$  denote the BS colocated with $x_0$. Define $R \triangleq \|x_0\|$.    
The downlink SIR of the typical user can be expressed as 
\begin{equation}
{\rm SIR} = \frac{h_0 R^{-\alpha}}{h_1 R^{-\alpha} + \sum_{x \in \Phis \setminus \{x_0\}} \left(h_{x,1}+h_{x,2}\right) \ell(x)}, 
\end{equation}
where $h_0, h_1, \{h_{x,1}\}, \{h_{x,2}\} \sim {\rm gamma}(m,\frac{1}{m})$ are independent  fading variables and $\Phis$ is the simple point process version of $\Phi$.  
Let $\tilde{I} \triangleq   \sum_{x \in \Phis \setminus \{x_0\}} \left(h_{x,1}+h_{x,2}\right) \ell(x)$. $\tilde{I}$ can be rewritten as 
\begin{align}
\tilde{I} &= \sum_{x \in \Phis} \sum_{y \in \Phis \setminus \{x\}}\left( \left(h_{y,1}+h_{y,2}\right) \|y\|^{-\alpha}\right) \mathbf{1}(\Phi(b(o,\|x\|)	=  0)),  
\end{align}
where   
$b(o, r)$ denotes the open disk of radius $r$ at $o$.

For the tail property of the SIR, we have the following theorem. 
\begin{theorem} For all stationary BS point processes, where every BS is colocated with another one and the typical user is served by one nearest BS, if the fading is Nakagami-$m$, then
	\begin{equation}
	\Ps(\theta) 
	\sim \theta^{-(m+\delta)} \frac{\lambda \pi  m^{-\delta}}{2(m+\delta) \left( \Gamma(m) \right)^2}\mathbb{E}_o^!(I_{\infty}^{-\delta}) \bigg(  \sum_{i=0}^m \binom{m}{i} \Gamma(m+i) \bar{\Gamma}(m-i)  \bigg) , \quad \theta \to \infty, 
	\label{main_asymp}
	\end{equation}
	where $\delta \triangleq 2/\alpha$, $I_{\infty} \triangleq \sum_{y \in \Phis} \left(h_{y,1}+h_{y,2}\right) \|y\|^{-\alpha}$,  and 
	\begin{equation}
	\bar{\Gamma}(m-i)=\left\{ \begin{array}{ll}
	\delta \Gamma(m-i+\delta), & i<m \\
	\Gamma(1+\delta), & i=m. 
	\end{array} \right.
	\end{equation}
	
	\label{MainResult}
\end{theorem}
\begin{IEEEproof}
	See Appendix \ref{sec:appA}. 
\end{IEEEproof}

By Theorem \ref{MainResult}, we know that in the log-log plot of $\Ps(\theta)$ v.s. $\theta$, the slope of  $\Ps(\theta)$ is $-(m+\delta)$, as $\theta \to \infty$. The slope only depends on the fading type and the path loss exponent.  

\subsubsection{Bounded path loss model}
\label{sec::asymp_nonsimple_bounded_infty_cellular}
The system model  is the same as that in Section \ref{sec::asymp_nonsimple_singular_infty_cellular}, except that $\Phis$ is assumed to be a uniform PPP with intensity $\lambda/2$, the fading is Rayleigh and $\ell(x) = (\epsilon+\|x\|^{\alpha})^{-1}$, where $\alpha > 2$ and  $\epsilon>0$.  
We have 
\begin{align}
P_{s,{\rm PPP}}(\theta) &=\mathbb{P} \left( \frac{h_0 \ell(R)}{h_1 \ell(R) +\tilde{I} }> \theta  \right) \nonumber\\
&= \mathbb{E}\left[e^{-h_1 \theta} \right] \mathbb{E}\left[  e^{-\theta \left(\epsilon+R^{\alpha} \right) \tilde{I}} \right]     \nonumber\\ 
&= \frac{1}{1+\theta} \mathbb{E}\left[  e^{-\theta \left(\epsilon+R^{\alpha} \right) \tilde{I}} \right]    . 
\end{align}

Using the same method as  in Section \ref{sec::asymp_simple_bounded_infty_cellular}, we obtain  that as $\theta \to \infty$, 
\begin{align}
P_{s,{\rm PPP}}(\theta) 
&\sim   \frac{1}{1+\theta}  \frac{\lambda \theta^{-\delta} }{2} \int_{\mathbb{R}^2} \exp\left(  -2 \pi \lambda  \left(\theta \epsilon+ \|x\|^{\alpha}\right)( \left(\theta \epsilon+ \|x\|^{\alpha}\right)+\epsilon)^{\delta-1}\int_0^{\infty} \frac{r^{\alpha}+\frac{1}{2} }{\left(  r^{\alpha} + 1 \right)^2} r\mathrm{d}r  \right)   \mathrm{d}x  \nonumber\\
&\sim   \frac{1}{1+\theta}  \frac{\lambda \theta^{-\delta} }{2}  \exp\left(- 2\pi \lambda \theta^{\delta} \epsilon^{\delta} \int_0^{\infty} \frac{r^{\alpha}+\frac{1}{2} }{\left(  r^{\alpha} + 1 \right)^2} r\mathrm{d}r \right)  \nonumber\\
&\quad \cdot \int_{\mathbb{R}^2} \exp\left(  -2 \pi \lambda \left( \left(\theta \epsilon+ \|x\|^{\alpha}\right)( \left(\theta \epsilon+ \|x\|^{\alpha}\right)+\epsilon)^{\delta-1}  -\theta^{\delta} \epsilon^{\delta} \right) \int_0^{\infty} \frac{r^{\alpha}+\frac{1}{2} }{\left(  r^{\alpha} + 1 \right)^2} r\mathrm{d}r  \right)   \mathrm{d}x.   
\end{align}

Thus, $\log(\Ps(\theta)) = \Theta( \theta^{\delta} )$, as $\theta \to \infty$.

\subsection{Discussion}
For non-trivial asymptotics of $\mathbb{P}(S/I>\theta)$ as $\theta\to \infty$, we need $0\in\supp(1/S)\cup\supp(I)$.
If this is not the case, there is a $\theta_0>0$ s.t.~$\Ps(\theta_0)=0$ $\forall \theta<\theta_0$, and thus the tail is trivial.

In simple  ad hoc models, for both the singular and bounded path loss models, 
the tail of the distribution of $S$  is 
determined only by the tail of the fading distribution;   the lower tail of the distribution of $I$ is determined  by both the lower tail of the fading distribution  and the tail of the nearest-interferer distance distribution. For Nakagami-$m$ fading, the distribution of the fading variable has an exponential tail and its lower tail follows a power law. If we fix the locations of all the interferers (i.e., conditioned on $\Phi$),   the lower tail of the distribution of $I$   decays faster than any power law, since $I$ is a sum of  a countable  number of weighted fading variables.  
If we fix all fading variables, the lower tail of the  distribution of $I$ depends on the tail of the distribution of the nearest-interferer distance.  
Note that we have not proved our results for all m.i. point processes. For the PPP,   the tail of the distribution of the nearest-interferer distance decays exponentially and thus  the lower tail of the distribution of $I$ is bounded by an exponential decay. 
So, $\Ps(\theta)$ decays faster than any power law---it is, in fact, proved to decay exponentially. 

In non-simple  ad hoc models, one interferer's location is deterministic. Thus, a necessary condition for $0\in\supp(I)$ is that 
$0\in\supp(h)$, where $h$ is a fading variable.
Similar to the simple ad hoc models, $\Ps(\theta)$ decays faster than any power law and is proved to decay exponentially.

In simple cellular models with singular path loss, whether $0\in\supp(1/S)$ is determined by the tail of the fading distribution and the lower tail of the distribution of  $\|x_0\|$, and  the lower tail of the distribution of $I$  decays faster than any power law. We  proved that $\Ps(\theta)$ decays by the power law and  the lower tail of the distribution of $\|x_0\|$ (and not the fading distribution) dominates the decay order, so the decay order is a function of $\delta$. 
In non-simple cellular models with singular   path loss, there is an interferer at the serving BS's location. We proved that  $\Ps(\theta)$ decays as a light power law, and both the lower tail of the distribution of $\|x_0\|$  and the ratio of two i.i.d .~fading variables determine the decay order, so the decay order is a function (the sum) of $\delta$ and the fading parameter.  
For both simple and non-simple cellular models with bounded path loss, whether $0\in\supp(1/S)$ is  determined only by the tail of the fading distribution, and  the lower tail of  the distribution of $I$ decays faster than any power law. We proved that $\Ps(\theta)$ decays exponentially.

\section{Impact of the Nearest Interferer on the Asymptotics}
\label{sec:ImpactNearest}
\subsection{Assumptions}
In this section, we study how the nearest interferer affects the asymptotics. In other words, we analyze the SIR asymptotics  if we only consider the interference from the nearest interferer with power $I_0$. 
In ad hoc networks, we assume the distance from the nearest interferer to the receiver at the origin is $R$ with PDF $f_{R}$ and $\mathbb{E}[R^t] < \infty$, for all $t>0$.  
In cellular networks, we assume  the distance from the nearest BS to the receiver at the origin  is $R_0$ with PDF $f_{R_0}$,  the distance from the nearest interferer to the receiver at the origin is $R$ with conditional PDF $f_{R\mid R_0}$ and $\mathbb{E}[R_0^t] < \infty$ and $\mathbb{E}[R^t \mid R_0] < \infty$, for all $t>0$.   
We define $\textrm{SIR}_0 \triangleq \frac{S}{I_0}$ and $P_0(\theta) \triangleq \mathbb{P}(\textrm{SIR}_0 > \theta)$. 

Table \ref{T_SIR0} summarizes the asymptotic properties of $P_0(\theta)$. The \shadetwo{\textrm{shading}} indicates that the results are the same as those in Table \ref{T_total}. Those entries indicate that the nearest interferer alone determines the decay order,  and other interferers may only affect the pre-constant.

\needspace{2cm}
\subsection{Lower Tail of the SIR Distribution}
\subsubsection{Simple ad hoc models}
\label{sec::adsimpleR0}
For the singular path loss model, the result has been proved in Proposition \ref{lemma1}.    
For the bounded path loss model, 
we have 
\begin{align}
1-P_0(\theta) &= \mathbb{P}\left( \frac{h_0 \ell(b)}{h \ell(R)}< \theta  \right) \nonumber\\
&= \mathbb{E}_{R}\mathbb{P}\left( h_0 < \theta h \ell(b)^{-1} \ell(R)\mid R \right)\nonumber\\
&= \mathbb{E}_{R}\mathbb{P}\left( h_0 < \theta h \ell(b)^{-1} (\epsilon +R^{\alpha})^{-1} \mid R \right).
\end{align}

As was stated after Theorem \ref{Thm_nonsimple_singular_0}, for Nakagami-$m$ fading, we have $\mathbb{P}\left(h_0 <  \theta h \right) = \Theta(\theta^{m})$ as $\theta \to 0$. Using the dominated convergence theorem and L'Hospital's rule, we can prove that  as $\theta \to 0$,
\begin{align}
1-P_0(\theta) = \Theta(\theta^{m}).
\end{align}

\begin{table}
	\small
	\caption{Asymptotic Properties of $P_0(\theta)$ (``Simple": simple point processes; ``Duplicated": duplicated-2-point point processes)}
	\begin{center}
		\begin{tabular}{|c|c|c|}
			\hline
			Models &$\theta \to 0$ &$\theta \to \infty$  \\ \hline 
			Simple \& Ad Hoc \& Singular path loss & \shadetwo{1-P_0(\theta) = \Theta( \theta^{\delta})} & $P_0(\theta)  = \Theta( \theta^{-m})$  \\ \hline
			Simple \& Ad Hoc \& Bounded path loss &\shadetwo{1-P_0(\theta) = \Theta(  \theta^{m})}  &   $P_0(\theta)  = \Theta( \theta^{-m})$  \\ \hhline{|=|=|=|}
			Simple \& Cellular \& Singular path loss  &\shadetwo{1-P_0(\theta) = \Theta( \theta^{m}) } & \shadetwo{P_0(\theta)  = \Theta( \theta^{-\delta})}  \\ \hline
			Simple \& Cellular \& Bounded path loss & \shadetwo{1-P_0(\theta) = \Theta( \theta^{m}) } &   $P_0(\theta)  = \Theta( \theta^{-m})$    \\ \hhline{|=|=|=|}
			Duplicated \& Ad Hoc \& Singular path loss  & \shadetwo{1-P_0(\theta) = \Theta( \theta^{\delta})} &    $P_0(\theta)  = \Theta( \theta^{-m})$  \\ \hline
			Duplicated \& Ad Hoc \& Bounded path loss &  \shadetwo{1-P_0(\theta) = \Theta( \theta^{m})}   &  $P_0(\theta)  = \Theta( \theta^{-m})$   \\ \hhline{|=|=|=|} 
			Duplicated \& Cellular \& Singular path loss & \shadetwo{1-P_0(\theta) = \Theta(  \theta^{m}) }  & $P_0(\theta)  = \Theta( \theta^{-m})$  \\ \hline
			Duplicated \& Cellular \& Bounded path loss & \shadetwo{1-P_0(\theta) = \Theta( \theta^{m})}   &   $P_0(\theta)  = \Theta( \theta^{-m})$  \\ \hline
		\end{tabular}
	\end{center}
	\label{T_SIR0}
\end{table}%

\subsubsection{Simple cellular models}
We can simply modify the proof of Theorem 1 in \cite{GuoTCOM15} and prove that $1-P_0(\theta) = \Theta(\theta^{m})$, as $\theta \to 0$.

\subsubsection{Non-simple ad hoc models}
Here the nearest interferer at most at distance $b$, since the point at $b$ is duplicated.
For the singular path loss model, we have 
\begin{align}
1-P_0(\theta) &= \mathbb{P}\left( \frac{h_0 \ell(b)}{h \ell(\min\{R,b\})}< \theta  \right) \nonumber\\
&=  \mathbb{P}(R<b) \mathbb{E}_{R}[\mathbb{P}\left( h_0 < \theta h \ell(b)^{-1} \ell(R)\mid R \right)\mid R<b ] + \mathbb{P}(R\geq b) \mathbb{E}_{R}[\mathbb{P}\left( h_0 < \theta h  \right) \mid R \geq b] \nonumber\\
&= \mathbb{P}(R<b) \mathbb{E}_{R}[\mathbb{P}\left( h_0 < \theta h \ell(b)^{-1} \ell(R)\mid R \right)\mid R<b ] + \mathbb{P}(R\geq b)  \mathbb{P}\left( h_0 < \theta h  \right).
\label{eq_adhoc}
\end{align}

From Section  \ref{sec::adsimpleR0}, we know that the first term in \eqref{eq_adhoc}  is $\Theta(\theta^{\delta})$ and the second term is $\Theta(\theta^{m})$. So, $1-P_0(\theta) = \Theta(\theta^{\delta})$, as $\theta \to 0$.  

For  the bounded path loss model, we can apply the same methods as in the corresponding cases in Section \ref{sec::adsimpleR0} and obtain the results in Table \ref{T_SIR0}. 

\subsubsection{Non-simple cellular models}
We have $R=R_0$. Thus, $\textrm{SIR}_0 = h_0/h_1$. We can easily obtain the results in Table \ref{T_SIR0}. 

We observe that in all cases, the lower tail is identical to the one when all interferers are considered. For the duplicated
point processes, this implies that duplicating only the nearest interferer (as is the case for edge users in cellular networks)
again results in the same scaling.

\subsection{Tail of the SIR Distribution}

\subsubsection{Simple ad hoc models}

For both singular and bounded path loss models, we have 
\begin{align}
P_0(\theta) &= \mathbb{P}\left( \frac{h_0 \ell(b)}{h \ell(R)}> \theta  \right) \nonumber\\
&= \mathbb{E}_{R}\mathbb{P}\left( h < \theta^{-1} h_0 \ell(b) \ell(R)^{-1} \mid R \right).
\end{align}

Since $\mathbb{E}[R^t] < \infty$, for all $t>0$, using the dominated convergence theorem and L'Hospital's rule, we obtain that $P_0(\theta) = \Theta(\theta^{-m})$, as $\theta \to \infty$.

\subsubsection{Simple cellular models}

For   the singular path loss model, we define $\bar{R}_2$ as the distance from the origin to its   nearest point of $\Phi_o^!$ and assume $\mathbb{E}(\bar{R}_2^2) < \infty$. We have
\begin{align}
P_0(\theta) &= \mathbb{P}\left( \frac{h_0 \ell(R_0)}{h \ell(R)}> \theta  \right) \nonumber\\
&= \mathbb{E} \sum_{x\in \Phi}   \bar{F}_{h_0} \left(  \theta  \|x\|^{\alpha} h  \|\bar{x}_2\|^{-\alpha}     \right) \mathbf{1}\left(  \Phi(b(o,\|x\|)) = 0 \right)\mathbf{1}\left(  \Phi(b(o,\|\bar{x}_2\|)) = 1 \right)\nonumber\\
&= \lambda \int_{\mathbb{R}^2}\mathbb{E}_o^ !  \bar{F}_{h_0} \left( \theta  \|x\|^{\alpha} h  \|\bar{x}_2\|^{-\alpha}    \right) \mathbf{1}\left(  b(o,\|x\|) \textrm{empty} \right) \mathbf{1}\left(  \Phi_x(b(o,\|\bar{x}_2\|)) \setminus \{x\}= 0 \right)  \mathrm{d}x\nonumber\\
&\stackrel{(a)}{=} \lambda \theta^{-\delta} \int_{\mathbb{R}^2}\mathbb{E}_o^!  \bar{F}_{h_0} \left(      \|x\|^{\alpha} h  \|\bar{x}_2\|^{-\alpha}      \right) \mathbf{1}\left(   b(o,\|x\|\theta^{-\delta/2}) \textrm{ empty} \right)  \nonumber\\
& \quad \cdot \mathbf{1}\left(   \Phi_{x\theta^{-\delta/2} }(b(o,\|\bar{x}_2\|)) \setminus \{x\theta^{-\delta/2}\}= 0 \right)  \mathrm{d}x\nonumber\\
&\stackrel{(b)}{\sim} \lambda \theta^{-\delta} \int_{\mathbb{R}^2}\mathbb{E}   \bar{F}_{h_0} \left(      \|x\|^{\alpha} h  \|\bar{x}_2\|^{-\alpha}   \right)   \mathbf{1}\left(   \Phi_{o}^!(b(o,\|\bar{x}_2\|))  = 0 \right)    \mathrm{d}x, \quad  \theta \to \infty\nonumber\\
&\stackrel{(c)}{=} \lambda \theta^{-\delta} \mathbb{E}\left[\left(h  \bar{R}_2^{-\alpha} \right)^{-\delta} \right] \int_{\mathbb{R}^2}    \bar{F}_{h_0} \left(     \|x\|^{\alpha}   \right)     \mathrm{d}x, \quad  \theta \to \infty \nonumber\\
&= \lambda \theta^{-\delta} \mathbb{E}\left[h^{-\delta} \right] \mathbb{E}\left[\bar{R}_2^2\right] \pi \mathbb{E}(h^{\delta}), \quad  \theta \to \infty  ,
\end{align}
where  $\Phi_x = \{y \in \Phi: y+x \} $ is a translated version of $\Phi$, $(a)$ follows by using the substitution $x\theta^{\delta/2} \to x$, $(b)$ follows since $ \mathbf{1}\left(   b(o,\|x\|\theta^{-\delta/2}) \textrm{ empty} \right) \to 1$, and $(c)$ follows by using the substitution $x(h  \|\bar{x}_2\|^{-\alpha})^{\delta/2} \to x$.
Thus,  $P_0(\theta) = \Theta(\theta^{-\delta})$, as $\theta \to \infty$. 

For  the bounded path loss model, we have 
\begin{align}
P_0(\theta) &= \mathbb{P}\left( \frac{h_0 \ell(R_0)}{h \ell(R)}> \theta  \right) \nonumber\\
&= \mathbb{E}_{R_0} \mathbb{E}_{R \mid R_0}\mathbb{P}\left( h < \theta^{-1} h_0 \ell(R_0) \ell(R)^{-1} \mid R, R_0 \right).
\end{align}

Since $\mathbb{E}[R_0^t] < \infty$ and $\mathbb{E}[R^t \mid R_0] < \infty$, for all $t>0$, using the dominated convergence theorem and L'Hospital's rule, we obtain that $P_0(\theta) = \Theta(\theta^{-m})$, as $\theta \to \infty$.

\subsubsection{Non-simple ad hoc models}

For both singular and bounded path loss models, we can apply the same methods as in the corresponding cases in Section \ref{sec::adsimpleR0} and obtain the results in Table \ref{T_SIR0}.

\subsubsection{Non-simple cellular models}
We have $R=R_0$. Thus, $\textrm{SIR}_0 = h_0/h_1$, and the results in Table \ref{T_SIR0} follow easily. 

\subsection{Discussion}
Since $I\geq I_0$, we have $\frac{S}{I_0} \geq \frac{S}{I} $, and thus $1-P_0(\theta) \leq 1-\Ps(\theta)$. Consequently, $\theta \to 0$, $1-P_0(\theta)$ decays faster than or in the same order as $1-\Ps(\theta)$; also, since  $P_0(\theta) \geq  \Ps(\theta) $, as $\theta \to \infty$, $P_0(\theta)$ decays slower than or in the same order as $\Ps(\theta)$. This is consistent with the results in Tables \ref{T_total} and \ref{T_SIR0}.

\subsubsection{$\theta \to 0$}
For non-trivial asymptotics, we need $0\in\supp(S)\cup\supp(1/I_0)$.
In ad hoc models, for both the singular and bounded path loss models, whether $0\in\supp(S)$ is only determined by the  distribution of the fading variable near $0$.
 For the singular path loss model, whether $0\in\supp(1/I_0)$ is determined by both the tail of the fading distribution  and the lower tail of the  nearest-interferer distance distribution, while for the bounded path loss model,  it is determined only by  the tail of the fading distribution. Thus, for ad hoc models with bounded path loss, the decay order is only determined by the fading parameter $m$.  For simple  ad hoc models with singular path loss, we  proved that  the lower tail of the nearest-interferer distance distribution, other than the fading distribution, dominates the decay order and thus the decay order is only determined by $\delta$.

In cellular models, $R_0$ can be arbitrarily small and it always holds that $R \geq R_0$.  
For simple point processes, we  proved that it is   the fading variable, not the path loss exponent $2/\delta$,
that determines the decay order. For non-simple point processes, since $R=R_0$,
the fading variable alone determines the decay order.

\subsubsection{$\theta \to \infty$}
For non-trivial asymptotics, we need $0\in\supp(1/S)\cup\supp(I_0)$.
In simple ad hoc models, for both the singular and bounded path loss models, whether $0\in\supp(1/S)$ is  determined only by the tail of the fading distribution and whether $0\in\supp(I_0)$ is determined by both the lower tail of the  fading distribution and the tail of the nearest-interferer distance distribution.
We have proved that the fading distribution, rather than the nearest-interferer distance distribution, dominates the decay order. 

The  non-simple ad hoc models have the same asymptotics as the simple ones, since only the tail of the nearest-interferer distance distribution 
may affect the asymptotic property and in the non-simple case, the nearest-interferer distance cannot be larger than $b$. 

In simple cellular models, for   the singular   path loss model, whether $0\in\supp(1/S)$ is  determined by the tail of the fading distribution and the lower tail of the distribution of $R_0$, while  $0\in\supp(I_0)$ is determined by both the lower tail of the fading distribution and the tail of the distribution of $R$.  We have proved that the lower tail of the distribution of $R_0/R$   
(especially, that of $R_0$), other than the fading distribution, dominates the decay order, so the decay order is a function of $\delta$. For the bounded path loss model, whether $0\in\supp(1/S)$ is  determined only by the tail of the  fading distribution, while $0\in\supp(I_0)$ is determined by both the lower tail of the  fading distribution and the tail of the distribution of $R$. We have proved that the fading distribution, rather than the distribution of $R$, dominates the decay order.

In non-simple cellular models, since $R=R_0$,   the fading distribution alone determines the decay order.  

\subsection{Three Conjectures}
Based on the insight obtained from the asymptotic results for the cases where the total interference is taken into account and
where only the nearest interferer is considered, we can formulate three conjectures that succinctly summarize and generalize
our findings for both cellular and ad hoc models---assuming they hold.

Let $x_0$ be the desired transmitter and $y$ the nearest interferer. 
\begin{conjecture}[Lower heavy tail]
\[ 0\in\supp(\ell(x_0)/\ell(y)) \quad \Longleftrightarrow \quad 1-\Ps(\theta) = \Theta(\theta^{\delta}),\quad\theta\to 0. \]
\end{conjecture}
In words, this conjecture states that if the (mean) power of the nearest interferer can get arbitrarily large relative to
that of the desired transmitter, then
the outage probability only decays with $\theta^{\delta}$ as $\theta\to 0$, and vice versa.
Conversely, if the ratio is bounded away from $0$,
the only reason why the SIR can get small is the fading, hence the fading statistics determine the scaling. 

\begin{conjecture} [Heavy tail]
\[ 0\in\supp(1/\ell(x_0)) \quad \Longleftrightarrow \quad \Ps(\theta)=\Theta(\theta^{-\delta}),\quad\theta\to\infty \]
\end{conjecture}
In words, if the (mean) signal power can get arbitrarily large, the success probability has a heavy tail with
exponent $\delta$. Conversely, if the signal power is bounded,
then the only reason why the SIR can get big is the fading, hence the fading will
(co)determine the scaling (and if only the nearest interferer is considered, the fading parameter $m$
alone determines the scaling).

\begin{conjecture}[Exponential tail] 
For Rayleigh fading, 
\[ 0\notin\supp(1/\ell(x_0))  \quad \Longleftrightarrow \quad \Ps(\theta)=e^{-\Theta(\theta^\delta)},\quad\theta\to\infty. \]
\end{conjecture}
In words, if the signal power is bounded and the fading is exponential, the tail will be exponential with 
$\log \Ps(\theta)=\Theta(\theta^\delta)$.
\section{Conclusions}
\label{conclusions}
We considered a comprehensive class of wireless network models where transmitters/BSs follow general point processes
and analyzed the asymptotic properties of the lower and upper tails of the SIR distribution. 

In all cases, the lower tail of the SIR distribution decays as a power law.
Only the path loss exponent or the fading parameter determines the asymptotic order for ad hoc networks; only the  fading parameter determines the asymptotic order for cellular networks. This indicates that we can use the SIR distribution of the PPP  to approximate the SIR distribution of  non-Poisson models in  the high-reliability regime by applying a horizontal shift, which can be obtained using the mean interference-to-signal ratio (MISR) defined in \cite{MISR_Ganti2015}. 
For the tail of the SIR distribution, only cellular network models with singular path loss have power-law tails. In these cases, we can approximate the tail of the SIR distribution  of  non-Poisson networks using the corresponding Poisson result and the expected fading-to-interference ratio (EFIR) defined in \cite{MISR_Ganti2015}. For cellular networks with bounded path loss and ad hoc networks, we mainly investigated the Poisson case with Rayleigh fading and showed that the tail of the SIR distribution decays exponentially. 
The exponential decay does not imply that it is not possible to approximate the tail of the SIR distribution for non-Poisson networks
using the distribution of Poisson networks.
We conjecture that the same approximation method as for polynomial tails can be applied, but a proof of the feasibility of this
approach is beyond the scope of this paper and left for future work.

Moreover, we  investigated the impact of the nearest interferer on the asymptotic properties of the SIR distribution. If the nearest interferer is the only interferer in the networks, we proved that the scaling of the lower tail of the SIR distribution remains the same as when all interferers are taken into account, which means that the nearest interferer alone determines the decay order.
In contrast, for the tail of the SIR distribution, the nearest interferer alone does not dominate the asymptotic trend, and the tail decays as a power law, with a light tail in most cases and a heavy tail for simple cellular networks with singular path loss.  These findings are
summarized and generalized in three conjectures.

\appendices
\section{Proof of Theorem \ref{Thm_nonsimple_singular_0}}
\label{sec:app_nonsimple_singular_0}
\begin{IEEEproof}
Using the same method as in the proof of Theorem 5.6 in \cite{NOW}, we can show that 
\begin{equation}
\mathbb{P}(\tilde{I} \geq y) \sim \frac{\pi \lambda}{2} \mathbb{E}[\left(h_a+h_b\right)^{\delta}] y^{-\delta}, \quad y \to \infty.
\label{heavy_tail_I_case2}
\end{equation}

The success probability can be rewritten as 
\begin{align}
\Ps(\theta) &= 1- \mathbb{P}\left(    \tilde{I} >h_0 b^{-\alpha}  \theta^{-1} - h_1 b^{-\alpha}   \right)  \nonumber\\
		&= 1-  \mathbb{E}_{h_0,h_1}\left[ \mathbb{P}\left(    \tilde{I}  >h_0 b^{-\alpha}  \theta^{-1} - h_1 b^{-\alpha} \right) \mathbf{1}(h_0 b^{-\alpha}  \theta^{-1} \geq h_1 b^{-\alpha})  \mid h_0, h_1   \right]     \nonumber\\
		&\quad -  \mathbb{E}_{h_0,h_1}\left[ \mathbb{P}\left(   \tilde{I} >h_0 b^{-\alpha}  \theta^{-1} - h_1 b^{-\alpha}  \right)  \mathbf{1}(h_0 b^{-\alpha}  \theta^{-1} < h_1 b^{-\alpha})   \mid h_0, h_1   \right]    \nonumber\\
		&= 1-  \mathbb{E}_{h_0,h_1}\left[ \mathbb{P}\left(    \tilde{I} >h_0 b^{-\alpha}  \theta^{-1} - h_1 b^{-\alpha} \right) \mathbf{1}(h_0 \theta^{-1} \geq h_1 )  \mid h_0, h_1   \right]     -  \mathbb{P}\left(   h_0   \theta^{-1} < h_1 \right)    .   
\end{align}
Thus, 
\begin{align}
\lim_{\theta \to 0} \frac{1- \Ps(\theta)}{\theta^{\delta}} &=\lim_{\theta \to 0} \frac{ \mathbb{E}_{h_0,h_1}\left[ \mathbb{P}\left(   \tilde{I}  >h_0 b^{-\alpha}  \theta^{-1} - h_1 b^{-\alpha} \right) \mathbf{1}(h_0 \theta^{-1} \geq h_1 )  \mid h_0, h_1   \right] +  \mathbb{P}\left(   h_0   \theta^{-1} < h_1 \right)  }{\theta^{\delta}} \nonumber\\
		&=   \lim_{\theta \to 0} \frac{ \mathbb{E}_{h_0,h_1}\left[ \mathbb{P}\left(   \tilde{I}  >h_0 b^{-\alpha}  \theta^{-1} - h_1 b^{-\alpha} \right) \mathbf{1}(h_0 \theta^{-1} \geq h_1 )  \mid h_0, h_1   \right]  }{\theta^{\delta}}  \nonumber\\ 
		&\stackrel{(a)}{=}    \mathbb{E}_{h_0,h_1}\left[  \lim_{\theta \to 0}  \frac{ \mathbb{P}\left(   \tilde{I}  >h_0 b^{-\alpha}  \theta^{-1} - h_1 b^{-\alpha} \right) \mathbf{1}(h_0 \theta^{-1} \geq h_1 )   }{\theta^{\delta}}  \mid h_0, h_1   \right]     \nonumber\\ 
		&\stackrel{(b)}{=}     \mathbb{E}_{h_0,h_1}\left[  \lim_{\theta \to 0} \frac{\pi \lambda b^2  \mathbb{E}[\left(h_a+h_b\right)^{\delta}] }{2}  \frac{   \left( h_0 \theta^{-1} - h_1  \right)^{-\delta}   \mathbf{1}(h_0 \theta^{-1} \geq h_1 )   }{\theta^{\delta}}  \mid h_0, h_1   \right]   \nonumber\\ 
		 &=  \frac{\pi \lambda b^2  \mathbb{E}[\left(h_a+h_b\right)^{\delta}] }{2}  \mathbb{E}_{h_0,h_1}\left[  \lim_{\theta \to 0} \left( h_0  - h_1 \theta \right)^{-\delta}   \mathbf{1}(h_0  \geq h_1 \theta)    \mid h_0, h_1   \right]      \nonumber\\ 
		 &= \frac{ \pi \lambda b^2}{2} \mathbb{E}[\left(h_a+h_b\right)^{\delta}] \mathbb{E}[h^{-\delta}],  
\end{align}
where $(a)$ follows from the dominated convergence theorem and $(b)$ follows from \eqref{heavy_tail_I_case2}. 

Next, we show that Nakagami-$m$ fading meets the fading constraints wherein $K=m$.
Let $h_a, h_b \in  \textrm{gamma}(m,1/m)$.  We have 
\begin{align}
\mathbb{E}[h_a^\delta]  = \int_0^{\infty} \frac{m^m}{\Gamma(m)} x^{m-1+\delta} e^{-mx} \mathrm{d}x = \frac{\Gamma(m+\delta)}{m^{\delta}\Gamma(m)},
\end{align}
\begin{align}
\mathbb{E}[h_a^{-\delta}]  = \int_0^{\infty} \frac{m^m}{\Gamma(m)} x^{m-1-\delta} e^{-mx} \mathrm{d}x = \frac{m^{\delta}\Gamma(m-\delta)}{\Gamma(m)},
\end{align}
and 
\begin{align}
\lim_{\theta \to 0} \frac{\mathbb{P}[h_a<\theta h_b] }{\theta^m} &= \lim_{\theta \to 0} \frac{\mathbb{E}_{h_b}\left[ \int_0^{\theta h_b} \frac{m^m}{\Gamma(m)} x^{m-1} e^{-mx} \mathrm{d}x    \right] }{\theta^m}  \nonumber \\
 &\stackrel{(a)}{=} \lim_{\theta \to 0} \frac{\mathbb{E}_{h_b}\left[   \frac{m^m}{\Gamma(m)} \theta^{m-1} h_b^{m} e^{-m\theta h_b}    \right] }{m \theta^{m-1}}  \nonumber \\
 &=  \frac{m^{m-1}}{\Gamma(m)}  \lim_{\theta \to 0}  \mathbb{E}_{h_b}\left[    h_b^{m} e^{-m\theta h_b}    \right]   \nonumber \\
&=  \frac{m^{m-1}}{\Gamma(m)}  \mathbb{E}_{h_b}\left[    h_b^{m}      \right]   \nonumber \\
 &=   \frac{\Gamma(2m)}{m \Gamma(m)^2},
\end{align}
where $(a)$ follows by L'Hospital's rule.

\end{IEEEproof}

\section{Proof of Theorem \ref{MainResult}}
\label{sec:appA}
\begin{IEEEproof}
The SIR distribution can  be expressed as 
\begin{align}
\Ps(\theta) 
&=   \mathbb{P}\left( \frac{h_0 R^{-\alpha}}{h_1 R^{-\alpha} +  \tilde{I}} > \theta \right) \nonumber \\
&=   \mathbb{P}\left(h_0 > \theta (h_1 + R^{\alpha} \tilde{I})   \right) \nonumber \\
&=   \mathbb{E} \left[ \bar{F}_{h}\left( \theta (h_1 + R^{\alpha}   \tilde{I}  \right)\right]. 
\end{align}

So, 
\begin{align}
&\lim_{\theta \to \infty} \frac{\Ps(\theta)}{ \theta^{-(m+\delta)}}  = \lim_{\theta \to \infty} \frac{\mathbb{E} \left[ \bar{F}_{h}\left( \theta \left(h_1 + R^{\alpha}   \tilde{I}  \right) \right) \right]}{ \theta^{-(m+\delta)}} 
=\lim_{\theta \to \infty}  \frac{ \frac{\partial}{ \partial \theta}  \mathbb{E} \left[  \bar{F}_{h}\left( \theta \left(h_1 + R^{\alpha}   \tilde{I} \right)  \right) \right]}{- (m + \delta) \theta^{-(m+\delta+1) } }  \nonumber \\
&= \lim_{\theta \to \infty}\mathbb{E} \left[  \frac{ \frac{\partial}{ \partial \theta} {F}_{h}\left( \theta \left(h_1 + R^{\alpha}   \tilde{I}  \right) \right) }{ (m + \delta) \theta^{-(m+\delta+1) } } \right] \nonumber \\   
&=\frac{1}{(m + \delta)\Gamma(m)} \lim_{\theta \to \infty}\mathbb{E} \left[  \theta^{2m+\delta } m^m \left(h_1 + R^{\alpha}   \tilde{I}  \right)^m     e^{-\theta m \left(h_1 + R^{\alpha}   \tilde{I}  \right) }
 \right] \nonumber \\
&= \frac{m^m}{(m + \delta)\Gamma(m)} \lim_{\theta \to \infty}\mathbb{E} \left[  \theta^{2m+\delta } \left(\sum_{i = 0}^{m} \binom{m}{i} h_1^{i} \left( R^{\alpha} \tilde{I}\right)^{m-i} \right)  e^{-\theta m \left(h_1 + R^{\alpha}   \tilde{I}  \right) }
 \right] \nonumber \\
&= \frac{m^m}{(m + \delta)\Gamma(m)} \sum_{i = 0}^{m} \binom{m}{i} \lim_{\theta \to \infty}\mathbb{E} \left[   \left(  \theta^{m+i }  h_1^{i}  e^{-\theta m  h_1}   \right)  \left(  \theta^{m-i+\delta } \left( R^{\alpha} \tilde{I}\right)^{m-i} e^{-\theta mR^{\alpha} \tilde{I}} \right)  \right] \nonumber \\
&= \frac{m^m}{(m + \delta)\Gamma(m)} \sum_{i = 0}^{m} \binom{m}{i} \lim_{\theta \to \infty}\mathbb{E} \left[     \theta^{m+i }  h_1^{i}  e^{-\theta m  h_1}  \right]  \mathbb{E} \left[    \theta^{m-i+\delta } \left( R^{\alpha} \tilde{I}\right)^{m-i} e^{-\theta mR^{\alpha} \tilde{I}}   \right]  
\label{Asymp_main}
\end{align}
where ${F}_{h}$ and $\bar{F}_{h}$ are respectively the CCDF and the CDF of $h$.  

In the following, we evaluate $\lim_{\theta \to \infty}\mathbb{E} \left[   \theta^{m+i }  h_1^{i}  e^{-\theta m  h_1}   \right]$ and $\lim_{\theta \to \infty} \mathbb{E} \left[     \theta^{m-i+\delta } \left( R^{\alpha} \tilde{I}\right)^{m-i} e^{-\theta mR^{\alpha} \tilde{I}}  \right]$, separately.   
For $\lim_{\theta \to \infty}\mathbb{E} \left[   \theta^{m+i }  h_1^{i}  e^{-\theta m  h_1}   \right]$, we have  
\begin{align}
 \lim_{\theta \to \infty}\mathbb{E} \left[   \theta^{m+i }  h_1^{i}  e^{-\theta m  h_1}   \right] 
 &= \lim_{\theta \to \infty} \int_0^{\infty}  \frac{m^m}{\Gamma(m)}  \left(  \theta^{m+i }  x^{i}  e^{-\theta m x} \right)  x^{m-1} e^{-mx} \mathrm{d}x  \nonumber\\
 &= \lim_{\theta \to \infty}   \frac{\Gamma(m+i)}{m^i \Gamma(m)} \left(\frac{\theta}{1+\theta}\right)^{m+i}  \int_0^{\infty}  \frac{\left(m \left(1+\theta\right) \right)^{m+i}}{\Gamma(m+i)}    x^{m+i-1} e^{-m(1+\theta)x} \mathrm{d}x  \nonumber\\
  &= \lim_{\theta \to \infty}   \frac{\Gamma(m+i)}{m^i \Gamma(m)} \left(\frac{\theta}{1+\theta}\right)^{m+i}   \nonumber\\
  &= \frac{\Gamma(m+i)}{m^i \Gamma(m)}. 
\label{Asymp_fading}
\end{align}

For $\lim_{\theta \to \infty} \mathbb{E} \left[     \theta^{m-i+\delta } \left( R^{\alpha} \tilde{I}\right)^{m-i} e^{-\theta mR^{\alpha} \tilde{I}}  \right]$, we discuss the cases when $i<m$ and when $i = m$, separately. 

When $i < m$, let $g_i \sim {\rm gamma}(m-i,1)$, and  we have 
\begin{align}
\lim_{\theta \to \infty} \mathbb{E} \left[     \theta^{m-i+\delta } \left( R^{\alpha} \tilde{I}\right)^{m-i} e^{-\theta mR^{\alpha} \tilde{I}}  \right] 
&= \lim_{\theta \to \infty}   \frac{\delta \Gamma(m-i)}{m^{m-i}}  \frac{ \frac{\partial}{ \partial \theta}  \mathbb{E} \left[ 1- \int_0^{\theta mR^{\alpha} \tilde{I} }  \frac{1}{\Gamma(m-i)} x^{m-i -1}  e^{-x }     \mathrm{d}x    \right]}{\frac{\partial}{ \partial \theta} \theta^{-\delta}}  \nonumber\\
&\stackrel{(a)}{=} \lim_{\theta \to \infty}   \frac{\delta \Gamma(m-i)}{m^{m-i}}  \frac{   \mathbb{E} \left[ 1- \int_0^{\theta mR^{\alpha} \tilde{I} }  \frac{1}{\Gamma(m-i)} x^{m-i -1}  e^{-x }     \mathrm{d}x    \right]}{\theta^{-\delta}}  \nonumber\\
&= \lim_{\theta \to \infty}   \frac{\delta \Gamma(m-i)}{m^{m-i}}  \frac{   \mathbb{E} \left[ \bar{F}_{g_i}(\theta mR^{\alpha} \tilde{I} )   \right]}{\theta^{-\delta}}, 
\label{Asymp_casei}  
\end{align}
where $(a)$ follows by applying the L'Hospital's rule reversely and $\bar{F}_{g_i}$ is the CCDF of ${g_i}$. 

Using the same method as in the proof of Theorem 4 in \cite{MISR_Ganti2015}, we obtain   
\begin{align}
& \lim_{\theta \to \infty} \frac{   \mathbb{E} \left[ \bar{F}_{g_i}\left(\theta mR^{\alpha} \tilde{I} \right)   \right]}{\theta^{-\delta}} \nonumber\\
&= \lim_{\theta \to \infty}  \theta^{\delta} 
\sum_{x \in \Phis} \bar{F}_{g_i}\left( \theta m \|x\|^{\alpha}  \left(\sum_{y \in \Phis \setminus \{x\}}\left( \left(h_{y,1}+h_{y,2}\right) \|y\|^{-\alpha}\right)\right)\right) \mathbf{1}(\Phi(b(o,\|x\|)	=  0))     \nonumber\\
&\stackrel{(b)}{=}  \lim_{\theta \to \infty}  \frac{\theta^{\delta} \lambda}{2} \int_{\mathbb{R}^2} \mathbb{E}_o^! \left[ 
\bar{F}_{g_i}\left( \theta m \|x\|^{\alpha}  \left(\sum_{y \in \Phi_x}\left( \left(h_{y,1}+h_{y,2}\right) \|y\|^{-\alpha}\right)\right)\right) \mathbf{1}\left(b(o,\|x\|) \; {\rm  empty} \right)  \right]  \mathrm{d}x \nonumber\\
&\stackrel{(c)}{=}  \lim_{\theta \to \infty}  \frac{\lambda}{2} \int\limits_{\mathbb{R}^2} \mathbb{E}_o^! \left[ 
\bar{F}_{g_i}\left(  m \|x\|^{\alpha}  \left(\sum_{y \in \Phi_{x\theta^{-\delta/2}}}\left( \left(h_{y,1}+h_{y,2}\right) \|y\|^{-\alpha}\right)\right)\right) \mathbf{1}\left(b(o,\|x\|\theta^{-\delta/2}) \; {\rm  empty} \right)  \right]  \mathrm{d}x \nonumber\\
&\stackrel{(d)}{=}  \frac{\lambda}{2} \int_{\mathbb{R}^2} \mathbb{E}_o^! \left[ 
\bar{F}_{g_i}\left(  m \|x\|^{\alpha} I_{\infty} \right)   \right]  \mathrm{d}x  \nonumber\\
&\stackrel{(e)}{=}     \frac{\lambda}{2} m^{-\delta} \mathbb{E}_o^! \left[  I_{\infty}^{-\delta} \right]    \int_{\mathbb{R}^2}
\bar{F}_{g_i}\left(   \|x\|^{\alpha}   \right)  \mathrm{d}x   \nonumber\\
&=    \frac{\lambda}{2} m^{-\delta} \mathbb{E}_o^! \left[  I_{\infty}^{-\delta} \right]   \pi \delta \int_0^{\infty}
r^{\delta - 1} \bar{F}_{g_i}\left(   r \right)  \mathrm{d}r   \nonumber\\
&=    \frac{\lambda}{2} \pi  m^{-\delta} \mathbb{E}_o^! \left[  I_{\infty}^{-\delta} \right]  \mathbb{E}[g_i^{\delta}] \nonumber\\
&=    \frac{\lambda \pi  m^{-\delta} \Gamma(m-i+\delta)}{2\Gamma(m-i)} \mathbb{E}_o^! \left[  I_{\infty}^{-\delta} \right],   
\label{Asymp_-delta}
\end{align}
where $(b)$ follows from the Campbell-Mecke theorem, $\Phi_x \triangleq \{ y \in \Phi: y+x\}$  is a translated version of $\Phi$, $(c)$ follows by using the substitution $x\theta^{\delta/2} \to x$, $(d)$ follows by the   dominated convergence theorem  and the fact that  $\theta^{-\delta/2} \to 0$ and thus $\mathbf{1}\left(b(o,\|x\|\theta^{-\delta/2}) \; {\rm  empty} \right) \to 1$, $(e)$ follows by using the substitution $x\left(mI_{\infty}\right)^{\delta/2} \to x$. 
So, by substituting \eqref{Asymp_-delta} into \eqref{Asymp_casei}, it yields that 
\begin{align}
\lim_{\theta \to \infty} \mathbb{E} \left[     \theta^{m-i+\delta } \left( R^{\alpha} \tilde{I}\right)^{m-i} e^{-\theta mR^{\alpha} \tilde{I}}  \right] 
=   \frac{\delta \lambda \pi  \Gamma(m-i)}{2m^{m-i+\delta}}    \mathbb{E}_o^! \left[  I_{\infty}^{-\delta} \right]  \mathbb{E}[g_i^{\delta}].
\label{Asymp_casei_final}
\end{align}

When $i = m$,  let $g_m \sim {\rm exp}(1)$, and  we have 
\begin{align}
\lim_{\theta \to \infty} \mathbb{E} \left[     \theta^{m-i+\delta } \left( R^{\alpha}\tilde{I} \right)^{m-i} e^{-\theta mR^{\alpha}\tilde{I} }  \right]  
&= \lim_{\theta \to \infty} \mathbb{E} \left[     \theta^{\delta }  e^{-\theta mR^{\alpha}\tilde{I} }  \right] \nonumber\\
&=    \lim_{\theta \to \infty}  \theta^{\delta } \mathbb{E} \left[  \bar{F}_{g_m}\left(\theta mR^{\alpha} \tilde{I} \right) \right] \nonumber\\
&\stackrel{(a)}{=}   \frac{\lambda}{2} \pi  m^{-\delta} \mathbb{E}_o^! \left[  I_{\infty}^{-\delta} \right]  \mathbb{E}[g_m^{\delta}]   \nonumber\\
&=   \frac{\lambda}{2} \pi  m^{-\delta} \Gamma(1+\delta) \mathbb{E}_o^! \left[  I_{\infty}^{-\delta} \right], 
\label{Asymp_case0}
\end{align}
where $(a)$ follows from \eqref{Asymp_-delta}. 

Substituting \eqref{Asymp_fading}, \eqref{Asymp_casei_final} and \eqref{Asymp_case0} into \eqref{Asymp_main}, we obtain \eqref{main_asymp}.
\end{IEEEproof}

 \bibliographystyle{IEEEtran} 

\begin{thebibliography}{10}
\providecommand{\url}[1]{#1}
\csname url@samestyle\endcsname
\providecommand{\newblock}{\relax}
\providecommand{\bibinfo}[2]{#2}
\providecommand{\BIBentrySTDinterwordspacing}{\spaceskip=0pt\relax}
\providecommand{\BIBentryALTinterwordstretchfactor}{4}
\providecommand{\BIBentryALTinterwordspacing}{\spaceskip=\fontdimen2\font plus
\BIBentryALTinterwordstretchfactor\fontdimen3\font minus
  \fontdimen4\font\relax}
\providecommand{\BIBforeignlanguage}[2]{{%
\expandafter\ifx\csname l@#1\endcsname\relax
\typeout{** WARNING: IEEEtran.bst: No hyphenation pattern has been}%
\typeout{** loaded for the language `#1'. Using the pattern for}%
\typeout{** the default language instead.}%
\else
\language=\csname l@#1\endcsname
\fi
#2}}
\providecommand{\BIBdecl}{\relax}
\BIBdecl

\bibitem{CodingGain_book}
S.~Lin and D.~J. Costello, \emph{Error Control Coding}, 2nd~ed.\hskip 1em plus
  0.5em minus 0.4em\relax Englewood Cliffs, NJ: Prentice-Hall, 2004.

\bibitem{Diversity_multiplexing_Tse2003}
L.~Zheng and D.~N.~C. Tse, ``Diversity and multiplexing: a fundamental tradeoff
  in multiple-antenna channels,'' \emph{IEEE Transactions on Information
  Theory}, vol.~49, no.~5, pp. 1073--1096, May 2003.

\bibitem{High-SIR_Transmission_Ganti2011}
R.~K. Ganti, J.~G. Andrews, and M.~Haenggi, ``High-{SIR} transmission capacity
  of wireless networks with general fading and node distribution,'' \emph{IEEE
  Transactions on Information Theory}, vol.~57, no.~5, pp. 3100--3116, May
  2011.

\bibitem{Haenggi_JSAC09}
M.~Haenggi, J.~G. Andrews, F.~Baccelli, O.~Dousse, and M.~Franceschetti,
  ``Stochastic geometry and random graphs for the analysis and design of
  wireless networks,'' \emph{IEEE Journal on Selected Areas in Communications},
  vol.~27, no.~7, pp. 1029--1046, Sep. 2009.

\bibitem{GuoTCOM15}
A.~Guo and M.~Haenggi, ``Asymptotic deployment gain: A simple approach to
  characterize the {SINR} distribution in general cellular networks,''
  \emph{IEEE Transactions on Communications}, vol.~63, no.~3, pp. 962--976,
  Mar. 2015.

\bibitem{MISR2014}
M.~Haenggi, ``The mean interference-to-signal ratio and its key role in
  cellular and amorphous networks,'' \emph{IEEE Wireless Communications
  Letters}, vol.~3, no.~6, pp. 597--600, Dec. 2014.

\bibitem{MISR_Ganti2015}
R.~K. Ganti and M.~Haenggi, ``Asymptotics and approximation of the {SIR}
  distribution in general cellular networks,'' \emph{IEEE Transactions on
  Wireless Communications}, vol.~15, no.~3, pp. 2130--2143, Mar. 2016.

\bibitem{net:Baccelli09jsac}
F.~Baccelli, B.~Blaszczyszyn, and P.~M{\"u}hlethaler, ``Stochastic analysis of
  spatial and opportunistic {Aloha},'' \emph{IEEE Journal on Selected Areas in
  Communications}, vol.~27, no.~7, pp. 1105--1119, Sep. 2009.

\bibitem{net:Zhang14tit}
X.~Zhang and M.~Haenggi, ``The performance of successive interference
  cancellation in random wireless networks,'' \emph{IEEE Transactions on
  Information Theory}, vol.~60, no.~10, pp. 6368--6388, Oct. 2014.

\bibitem{Nigam2014}
G.~Nigam, P.~Minero, and M.~Haenggi, ``Coordinated multipoint joint
  transmission in heterogeneous networks,'' \emph{IEEE Transactions on
  Communications}, vol.~62, no.~11, pp. 4134--4146, Nov. 2014.

\bibitem{net:Nigam15jsac}
------, ``Spatiotemporal cooperation in heterogeneous cellular networks,''
  \emph{IEEE Journal on Selected Areas in Communications}, vol.~33, no.~6, pp.
  1253--1265, Jun. 2015.

\bibitem{ICIC_ICD_Xinchen2014}
X.~Zhang and M.~Haenggi, ``A stochastic geometry analysis of inter-cell
  interference coordination and intra-cell diversity,'' \emph{IEEE Transactions
  on Wireless Communications}, vol.~13, no.~12, pp. 6655--6669, Dec. 2014.

\bibitem{net:Mukherjee12jsac}
S.~Mukherjee, ``Distribution of downlink {SINR} in heterogeneous cellular
  networks,'' \emph{IEEE Journal on Selected Areas in Communications}, vol.~30,
  no.~3, pp. 575--585, Apr. 2012.

\bibitem{net:Blaszczyszyn15tit}
B.~Blaszczyszyn and H.~P. Keeler, ``Studying the {SINR} process of the typical
  user in {Poisson} networks by using its factorial moment measures,''
  \emph{IEEE Transactions on Information Theory}, vol.~61, no.~12, pp.
  6774--6794, Dec. 2015.

\bibitem{net:Miyoshi14aap}
N.~Miyoshi and T.~Shirai, ``A cellular network model with {Ginibre} configured
  base stations,'' \emph{Advances in Applied Probability}, vol.~46, no.~3, pp.
  832--845, Sep. 2014.

\bibitem{TON11}
R.~Giacomelli, R.~K. Ganti, and M.~Haenggi, ``Outage probability of general ad
  hoc networks in the high-reliability regime,'' \emph{IEEE/ACM Transactions on
  Networking}, vol.~19, no.~4, pp. 1151--1163, Aug. 2011.

\bibitem{NOW}
M.~Haenggi and R.~K. Ganti, ``Interference in large wireless networks,''
  \emph{Foundations and Trends in Networking}, vol.~3, no.~2, pp. 127--248,
  2009.

\bibitem{book}
M.~Haenggi, \emph{Stochastic Geometry for Wireless Networks}.\hskip 1em plus
  0.5em minus 0.4em\relax Cambridge University Press, 2012.

\bibitem{net:Guo16tcom}
A.~Guo, Y.~Zhong, M.~Haenggi, and W.~Zhang, ``{The Gauss-Poisson Process for
  Wireless Networks and the Benefits of Cooperation},'' \emph{IEEE Transactions
  on Communications}, vol.~64, no.~7, pp. 2985--2998, Jul. 2016.

\end{thebibliography}

\end{document}